# Compatibility of Zr$_2$AlC MAX phase-based ceramics with oxygen-poor, static liquid lead-bismuth eutectic

*Bensu Tunca[a,b,*], Thomas Lapauw[a,b], Carolien Callaert[c], Joke Hadermann[c], Remi Delville[a], El'ad N. Caspi[d], Martin Dahlqvist[e], Johanna Rosén[e], Amalraj Marshal[f], Konda G. Pradeep[f], Jochen M. Schneider[f], Jozef Vleugels[b] and Konstantina Lambrinou[a,g]*

[a] *SCK•CEN, Boeretang 200, 2400 Mol, Belgium*

[b] *KU Leuven, Department of Materials Engineering, Kasteelpark Arenberg 44, 3001 Leuven, Belgium*

[c] *University of Antwerp, EMAT, Groenenborgerlaan 171, 2020 Antwerp, Belgium*

[d] *Physics Department, Nuclear Research Centre-Negev, P.O. Box 9001, 84190 Beer Sheva, Israel*

[e] *Department of Physics, Chemistry and Biology (IFM), Linköping University, SE-581 83 Linköping, Sweden*

[f] *Materials Chemistry, RWTH Aachen University, 52074 Aachen, Germany*

[g] *University of Huddersfield, School of Computing and Engineering, Queensgate, Huddersfield HD1 3DH, UK*

[*] Corresponding author at: Boeretang 200, 2400 Mol, Belgium

E-mail address: bensu.tunca@kuleuven.be (B. Tunca)

## Abstract

This work investigates the compatibility of Zr$_2$AlC MAX phase-based ceramics with liquid LBE, and proposes a mechanism to explain the observed local Zr$_2$AlC/LBE interaction. The ceramics were exposed to oxygen-poor ($C_O \leq 2.2 \times 10^{-10}$ mass%), static liquid LBE at 500°C for 1000 h. A new Zr$_2$(Al,Bi,Pb)C MAX phase solid solution formed in-situ in the LBE-affected Zr$_2$AlC grains. Out-of-plane ordering was favorable in the new solid solution, whereby *A*-layers with high and low-Bi/Pb contents alternated in the crystal structure, in agreement with first-principles calculations. Bulk Zr$_2$(Al,Bi,Pb)C was synthesized by reactive hot pressing to study the crystal structure of the solid solution by neutron diffraction.

## 1. Introduction

The performance of most candidate structural and fuel cladding steels for Gen-IV lead-cooled fast reactors (LFRs) may degrade as result of their contact with the primary heavy liquid metal (HLM) coolant, which is either lead (Pb) or the lead-bismuth eutectic (LBE) alloy. An example of an accelerator-driven system (ADS) that uses liquid LBE as primary coolant and spallation target is the flexible fast-spectrum irradiation facility MYRRHA (multi-purpose hybrid research reactor for high-tech applications) currently under development at SCK•CEN, Belgium [1]. The inherent corrosiveness of HLM coolants results either in the liquid metal embrittlement of ferritic/martensitic steels [2] or in the dissolution corrosion of austenitic stainless steels [3,4]. The former is perceived as an appreciable reduction in steel ductility and the latter as the loss of steel alloying elements (e.g., Ni, Cr, Mn) into the HLM coolant. Liquid metal corrosion mitigation strategies for Gen-IV LFRs typically aim at reactor operation within well-defined conditions of temperature, HLM oxygen concentration, $C_O$, and component-specific lifetime [3] For particular applications, an alternative corrosion mitigation strategy consists in the development of new HLM-compatible materials, such as the MAX phase-based ceramics already tested in liquid Pb and LBE, showing promising first results [5,6]. The MAX phases are nanolaminated ternary carbides/nitrides with the $M_{n+1}AX_n$ general stoichiometry, where *M* is an early transition metal, *A* is an A-group element (typically groups 13-15) and *X* is C or N [7]. In general, the MAX phases are good thermal and electrical conductors, tolerant to mechanical damage, and machineable with conventional tools [7]. For nuclear applications close to the reactor core (e.g., nuclear fuel cladding materials), MAX phase nitrides are avoided, so as not to produce highly active nuclear waste by the formation of the long-lived isotope $^{14}C$ that results from the irradiation of N. Moreover, MAX phase compounds, such as $Ti_2AlC$ and $Ti_3SiC_2$, show good compatibility with HLMs [4,5] and promising neutron radiation tolerance above 600°C [8–12]; both properties make this group of ceramic materials appealing for fuel cladding applications in Gen-IV LFRs.

Earlier studies reported that the exposure of $Ti_3SiC_2$ and $Ti_2AlC$ MAX phase-based ceramics to oxygen-containing ($C_O \approx 10^{-6} - 10^{-8}$ mass%) liquid LBE and Pb formed μm-thick $TiO_2$ (rutile) and mixed $TiO_2/Al_2O_3$ oxide scales or dissolved 'parasitic' phases, such as TiAl in oxygen-poor ($C_O < 10^{-8}$ mass %) liquid LBE [5,6]. The reported test results have been summarized in a recent work by Lapauw et al. [13]. In the same work, 11 different MAX phase ceramics were exposed to oxygen-poor ($C_O \leq 2.2\times10^{-10}$ mass%), static LBE at 500°C for 1000 h. The tested MAX phases were the commercial grades Maxthal® 312 (nominally, $Ti_3SiC_2$) and Maxthal® 211 (nominally, $Ti_2AlC$) alongside various lab grades (i.e., $(Ti,Nb)_2AlC$, $Ti_3AlC_2$, $Nb_2AlC$, $(Zr,Ti)_{n+1}AlC_n$, $Ti_2SnC$, $(Nb_{0.85},Zr_{0.15})_4AlC_3$, $Zr_2AlC$, and $Zr_3AlC_2$). The latter three MAX phases were also exposed to oxygen-poor ($C_O \approx 5\times10^{-9}$ mass%), fast-flowing ($v \approx 8$ m/s) LBE at 500°C for 1000 h [13]. The majority of the MAX phases showed no interaction with LBE, except for the Zr-rich MAX phase ceramics that exhibited a local interaction. This work focuses on the detailed investigation of the interaction mechanism between the $Zr_2AlC$ MAX phase-based ceramic and oxygen-poor ($C_O \leq 2.2\times10^{-10}$ mass%), static liquid LBE after an exposure of 1000 h at 500°C.

## 2. Materials and methods

*2.1. Synthesis and LBE exposure conditions*

The synthesis route of $Zr_2AlC$ MAX phase-based ceramics was reported elsewhere [14]. In brief, fully dense $Zr_2AlC$ ceramics have been synthesized by reactive hot pressing of $ZrH_2$, Al and C powders (Zr:Al:C ≈ 2:0.8:1.2 or 2:1.1:0.95). XRD analysis of the $Zr_2AlC$ ceramics showed that they comprised 67 wt% $Zr_2AlC$ and 33 wt% ZrC [14]. The LBE exposure of a mirror-polished sample of the $Zr_2AlC$ ceramics was done in oxygen-poor ($C_O \leq 2.2\times10^{-10}$ mass%), static liquid LBE at 500°C for 1000 h [13]. The LBE $C_o$ was maintained low during testing by the constant supply of a reducing conditioning gas made of 95% Ar and 5% $H_2$ ($O_2$ < 10 ppm, $H_2O$ < 10 ppm) [3]; moreover, an electrochemical oxygen sensor ($Bi/Bi_2O_3$ reference electrode [15]) was used to monitor the LBE $C_o$ throughout the test. In order to study only the effect of the thermal annealing component of the LBE exposure, mirror-polished $Zr_2AlC$ bulk specimens were encapsulated in a quartz tube under vacuum (~$10^{-2}$ mbar) and were annealed at 500°C for 1000 h (i.e., at the exact same conditions of the LBE exposure) in a box furnace.

As the characterization of the LBE-affected $Zr_2AlC$ grains revealed the local formation of a new $Zr_2(Al,Bi,Pb)C$ MAX phase solution [13], it was decided to synthesize this new MAX phase solid solution by reactive hot pressing for detailed structural characterization. For this purpose, $ZrH_2$ (grain size <6 μm, >99% purity, Chemetall, Germany), LBE (44.5 wt% Pb and 55.5 wt% Bi, MCP 124, 5N Plus Lübeck GmbH, Belgium), Al (<5 μm, >99% purity, AEE, USA), and C (<5 μm, >99% purity, Asbury Graphite Mills, USA) powders were mixed in Zr:Al:LBE:C ≈ 2:0.55:0.55:0.95. The LBE powder was obtained by drilling in a block of as-cast LBE. The hot pressing temperature was 1250°C, while the other processing parameters (dwell time: 30 min, pressure: 30 MPa) were identical with those reported for the synthesis of $Zr_2AlC$-based ceramics [14].

*2.2. Materials characterization*

X-Ray diffraction (XRD) data were collected in the 5-75° $2\theta$ range with a step size of 0.02° and a counting time of 2 s per step, using a Cu $K_\alpha$ source in a Bragg-Brentano geometry operated at 40 kV and 40 mA (Seifert 3003) or 30 kV and 10 mA (Bruker D2 Phaser, Bruker). Metallographic cross-sections of the bulk ceramics were examined by scanning electron microscopy (SEM; Nova 600 NanoLab Dual Beam SEM/FIB, FEI) equipped with energy dispersive X-ray spectroscopy (EDS; EDAX) and electron backscattered diffraction (EBSD; Hikari XP EBSD camera, EDAX) for elemental and crystal orientation/phase mapping, respectively. The TEAM software (TEAM™ EDS Analysis System, EDAX) was used for EDS elemental mapping. Tips for atom probe tomography (APT) characterization of LBE-affected $Zr_2AlC$ grains were prepared in a FEI Helios NanoLab 660 dual beam microscope, by following a standard lift-out procedure [16,17]. Mild (5 kV) Ga ion milling was performed in the end to reduce Ga ion implantation and high-voltage milling artifacts [18]. The APT data were collected at 60 K tip temperature in laser mode, using a 4000X HR local electrode atom probe, and applying 30 pJ laser energy at 250 kHz pulse frequency [19,20]. The acquired data were reconstructed and analysed, using the IVAS 3.8.0 software provided by Cameca instruments.

Samples for transmission electron microscopy (TEM) examination were prepared by FIB (Nova NanoLab 600 DualBeam, FEI), using 30 kV Ga ions for milling and 5 kV Ga ions for cleaning. Selected area electron diffraction (SAED) patterns and bright field (BF) images were collected using TEM (JEOL ARM200), whereas high-resolution scanning TEM (HRSTEM) imaging was performed on an aberration-corrected Thermo Fisher Titan TEM (at 200 kV and 300 kV), equipped with a Super-X four detector EDS system and using the high-angle annular dark field (HAADF) and annular bright field (ABF) detectors. HAADF images are formed using high-angle, incoherently scattered electrons from the nucleus of atoms, the observed contrast is approximately proportional with $Z^2$ (where Z = atomic number) and, thus, heavy atoms appear brighter. On the other hand, ABF imaging uses the circumference of the transmitted beam, and both light and heavy atoms appear with comparable grey levels, which also allows locating the light atoms, but hinders the direct chemical identification of the different columns. For the HAADF-STEM simulations, the QSTEM [21] software was used.

The reactive hot pressed $Zr_2(Al,Bi,Pb)C$ ceramic was examined using SEM/EDS, while half of the produced disc was pulverized for XRD and neutron powder diffraction (NPD) analysis. NPD was performed at room temperature (RT) on the KARL powder double-axis diffractometer in the Israel Research Reactor No. 1 (IRR-1) (Nuclear Research Center, Soreq, Israel) [22]. The incident neutron beam ($\lambda$=0.982(1) Å) was obtained from the (220) reflection of a Cu single crystal monochromator. An angular step of 0.05° was used for a $2\theta$ scan in the 3.75-101.5° range. ~5 g of fine powder were loaded in a vanadium cylindrical sample holder ($\varnothing$ 10 mm, ~0.2 mm wall thickness). XRD and NPD data were analyzed with the Rietveld refinement method [23], using the MAUD [24] and FULLPROF [25] software, respectively. Rietveld analysis of the hot pressed $Zr_2(Al,Bi,Pb)C$ bulk ceramic was carried out by the simultaneous refinement of XRD and NPD results. Opting for joint refinement combined the advantages of XRD (high sensitivity to Pb/Bi and reliable statistics) with those of NPD (high sensitivity to C and ability to observe light elements, such as Al and C, in the vicinity of heavy ones, such as Pb and Bi). The refined structural parameters of the hot pressed $Zr_2(Al,Bi,Pb)C$ ceramic, as deduced from the combined NPD and XRD analysis, included the unit cell parameters and scale factors for all phases; the atomic position $z$ in the MAX phase; the isotropic atomic thermal displacement for each site in the two major phases; the Al and LBE occupancies in the MAX phase; the C occupancy in the two major phases; and a Lorentzian grain size for all phases.

*2.3. DFT calculations*

The in-situ formation of the $Zr_2(Al,Bi,Pb)C$ MAX phase solid solution in the LBE-affected areas, as confirmed by SEM, TEM and APT, inspired a series of first-principles calculations. Density functional theory (DFT) calculations were performed based on the projector augmented wave method [26,27], as implemented within the Vienna ab-initio simulation package (VASP) 5.4.1 [28–30]. The purpose of these calculations was to explore the partial substitution of Al by Bi and Pb. The used constraints were the assumptions of (a) a disordered solid solution of Al, Bi, Pb on the *A*-site in $Zr_2AlC$, (b) a fully populated *A*-lattice, i.e., no vacancies, (c) a fixed Bi:Pb ratio of 2:1 due to the commonly observed experimental Bi:Pb ratio in the LBE-affected areas, and (d) an Al-content in the 25–100% range. Details on the DFT calculations are provided in the supplementary information.

**3. Results**

*3.1. $Zr_2AlC$/LBE interaction: SEM/EDS and APT data*

The pristine (non-exposed) $Zr_2AlC$-based ceramic consisted of about 67 wt% $Zr_2AlC$ and 33 wt% ZrC and trace amount of Al-Zr intermetallic compounds between ZrC grains [14]. The Zr:Al:C atomic ratios in the $Zr_2AlC$ MAX phase were determined by X-ray photoelectron spectroscopy (XPS) to be 52.1:23.1:24.8 [14]. These ratios are consistent with the $Zr_2AlC$ stoichiometry (ideally, 50:25:25), with a slight abundance in Zr and deficiency in Al. The same authors reported NPD results where the C-site was determined to be fully occupied in both the $Zr_2AlC$ MAX phase and ZrC. Fig. 1 shows EBSD orientation and phase maps of the pristine $Zr_2AlC$ ceramic used in this work, where the determined phase assembly agrees with the previously reported data [14].

After the LBE exposure, XRD analysis was performed on cross-sectioned bulk samples to identify the microstructural response of the $Zr_2AlC$-based ceramic to the rather prolonged LBE exposure. Due to the XRD detection limit of 2-3 vol%, detection of the localised/shallow (sub-surface) interaction with LBE was not possible. No significant changes were identified in the LBE-exposed $Zr_2AlC$-based ceramic (see supplementary information for XRD patterns in Fig. S1 and for the lattice parameters of the constituent $Zr_2AlC$ and ZrC phases in Table S1). For $Zr_2AlC$, $a$ (0.2%) and $c$ (0.1%) increased slightly, while for ZrC, the $a$ value deviated only by -0.1% from the value reported in literature ($a$ = 4.686(1) Å [31]) and by +0.1% from the value reported upon $Zr_2AlC$ synthesis [14]. The phase assembly in the exposed ceramic did not change, as Rietveld refinement showed 66 (1) wt% $Zr_2AlC$ and 34 (1) wt% ZrC, which was the same as the phase assembly in the pristine ceramic (Fig. 1).

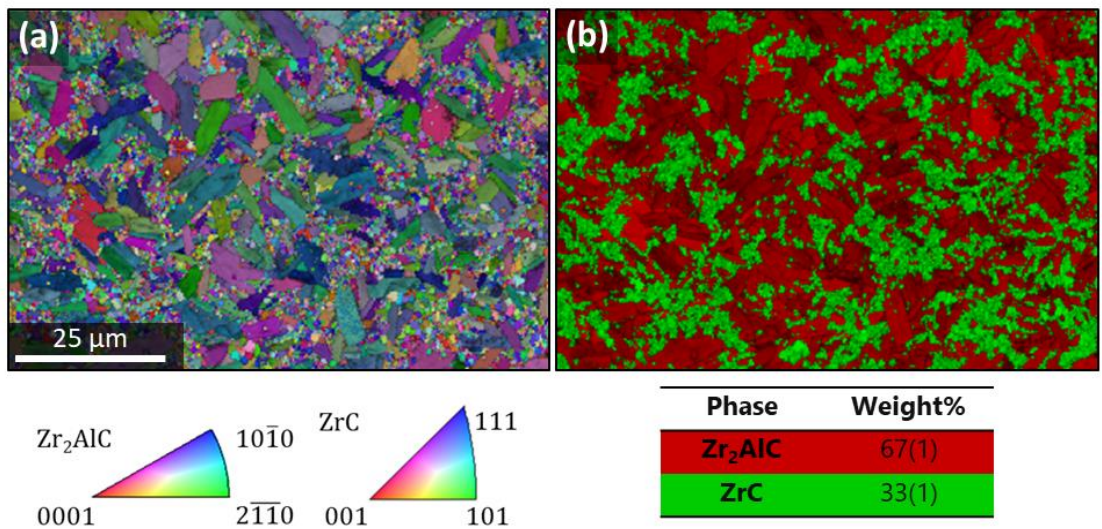

**Fig. 1.** (a) EBSD orientation map and (b) phase map of the pristine (as-sintered) $Zr_2AlC$ ceramic. This ceramic consists of the $Zr_2AlC$ MAX phase (red) and the ZrC competing phase (green).

The main features of the $Zr_2AlC$/LBE interaction are summarised in the SEM images of Fig. 2. The sample was kept submerged in the LBE bath by attaching it on a graphite plate with Mo wires [13]. One of its facets was polished prior to the LBE exposure, so as to facilitate the post-exposure investigation of the liquid metal corrosion effects [13]. Fig. 2 shows images of the post-exposure microstructure of the $Zr_2AlC$-based ceramic close to its polished facet. First, it is evident that some areas did not interact with liquid LBE (darker zones in Fig. 2a). Second, some regions suffered local LBE attack (brighter zones in Fig. 2a), which resulted in the in-situ formation of a MAX phase solid solution with the $Zr_2(Al,Pb,Bi)C$ general stoichiometry and Zr:(Al+Bi+Pb) ≈ 2, as determined by SEM/EDS point analyses [13]. EDS point analysis of various LBE-affected regions indicated a $Zr_2(Al_{0.50-0.79},Bi_{0.14-0.30},Pb_{0.08-0.18})C_y$ overall stoichiometry with Al:Bi:Pb ratios varying between 3:2:1 and 6:2:1. The total 'A' element (i.e., A = Al+Bi+Pb) content varied, resulting in a Zr:$A$ ratio that fluctuated between 2:0.9 and 2:1.15, while the ratio in $Zr_2AlC$ is 2:1. This means that both sub- and over-stoichiometric areas with respect to the $A$-element MAX phase content are present in the LBE-affected regions. Since the EDS analysis of the C content was not very accurate, the generic 'y' notation will be used herein to indicate the C stoichiometry. Fig. 2b shows the EDS phase map of an LBE-affected region with average $Zr_2(Al_{0.66},Bi_{0.22},Pb_{0.12})C_y$ composition, next to an unaffected part of the $Zr_2AlC$-based ceramic with $Zr_2Al_{1.33}C_y$ composition (see supplementary information Fig. S2 for individual elemental maps). The Al-overstoichiometric region at the LBE-interaction front is likely to be a build-up of Al diffusing out of the LBE-affected $Zr_2AlC$ grains, due to its partial substitution by the inward diffusing Bi and Pb. Fig. 2c shows a fully LBE-attacked area containing the new $Zr_2(Al,Bi,Pb)C$ MAX phase solid solution (brighter grains) and ZrC (darker grains). Fig. 2d shows the LBE interaction front, which is considered as the 'transition zone' between the fully LBE-affected and unaffected $Zr_2AlC$. Both 'parent' $Zr_2AlC$ and $Zr_2(Al,Bi,Pb)C$ MAX phase grains are observed in this zone, again in co-existence with the ZrC competing phase. The key feature of the transition zone are the 'stripy' grains indicating that the LBE attacked the $Zr_2AlC$ grains in a layer-by-layer manner; one such grain is pinpointed by a red arrow in Fig. 2d. Further APT and TEM examinations were performed in the transition zone, allowing the study of various features in both LBE-affected and unaffected $Zr_2AlC$ grains. APT and TEM data were collected from the following regions in Fig. 2d: (a) fully affected by LBE grains (red-dotted frame), (b) partially affected by LBE grains (yellow-dashed frame), and (c) unaffected by LBE grains (green-solid frame). The microstructure of the pristine (unaffected) $Zr_2AlC$-based bulk ceramic is shown in Fig. 2e.

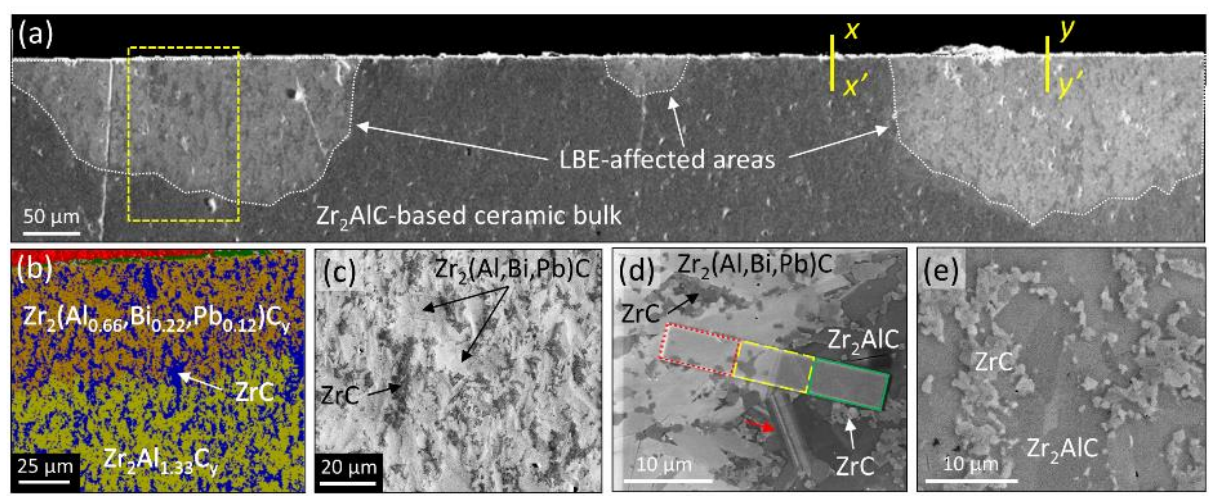

**Fig. 2.** SEM images of the polished facet of the as-exposed $Zr_2AlC$ ceramic: (a) overview showing local LBE interaction. (b) EDS phase map of the framed area in Fig. 2a. Images of the $Zr_2AlC$/LBE interaction: (c) fully affected zone, (d) transition zone, and (e) unaffected sample bulk. In Fig. 2d, the red arrow indicates basal LBE ingress, and the colored frames refer to areas studied by APT and TEM.

A long time after the end of the LBE exposure (several months to 1 year), an additional interaction was observed. This belated interaction with LBE manifested itself as delaminations along the basal planes within the LBE-affected $Zr_2AlC$ MAX phase grains mainly at the sample corners, but also in the thin foils prepared by FIB milling. Fig. 3a shows the sample corner with the deepest (~450 μm locally) LBE attack. The EDS phase mapping of this area (Fig. 3b) revealed 3 compositionally different areas: (a) an over-stoichiometric in Al area just ahead of the LBE interaction front ($Zr_2Al_{1.33}C_y$, blue area), (b) a $Zr_2(Al,Bi,Pb)C$ MAX phase solid solution area at the forefront of the LBE-affected zone ($Zr_2(Al_{0.73},Bi_{0.15},Pb_{0.12})C_y$, yellow area), and (c) an area deficient in 'A' elements, thus detected as mostly binary carbide ($ZrC_y$, red area) with occasional $Zr_2(Al,Bi,Pb)C$ grains at the outskirts of the LBE-affected zone (see supplementary information Fig. S2 for individual elemental maps).

The delaminations were observed in the $ZrC_y$-based area, as shown in the higher magnification image of Fig. 3c. In Fig. 3c, point-I corresponds to the new 211 MAX phase solid solution with $Zr_2(Al_{0.73},Bi_{0.15},Pb_{0.12})C_y$ composition, point-II to a delaminated MAX phase grain with ~$Zr_2Al_{0.4}C_y$ composition, and point-III to a $ZrC_y$ grain. All delaminated MAX phase grains had trace amounts of Bi and Pb, even though they were located in the LBE-affected zones. Similar delaminations, which occurred after the prolonged sample storage at RT, compromised the integrity of many TEM thin foils, leading to their untimely collapse. Fig. 3d shows a thin foil with a basal delamination in an LBE-attacked $Zr_2AlC$ MAX phase grain; this delamination was not present in the 'fresh' thin foil. EDS elemental mapping of an area around the delamination showed the presence of LBE between the separated basal planes (Fig. 3e).

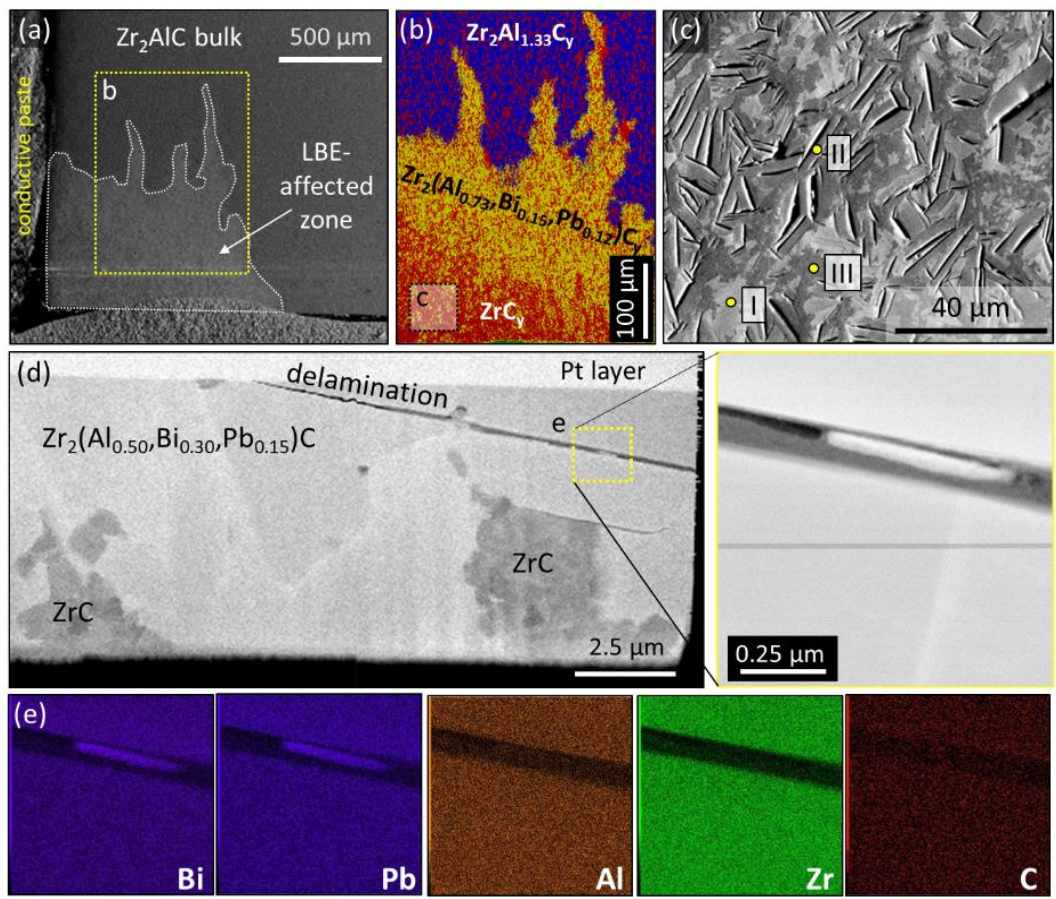

**Fig. 3.** (a) SEM image of an LBE-affected sample corner 1 year after the LBE exposure. (b) EDS phase map of the framed area in Fig. 3a. (c) SEM image of the framed area in Fig. 3b, showing many delaminations. (d) STEM image of a thin foil, showing a basal delamination in an LBE-affected $Zr_2AlC$ MAX phase grain. (e) STEM-EDS elemental mapping of the area around the basal delamination.

The LBE-affected regions were characterized by the outward diffusion of Al, which can be attributed to the fact that part of the Al atoms in $Zr_2AlC$ were pushed out by the incoming Bi and Pb atoms. The EDS elemental maps in Figs. 4b-4c show the presence of excess Al in the LBE-affected $Zr_2AlC$ MAX phase grain boundaries (GBs), along with Bi and Pb. The TEM examination of a cluster of finer ZrC grains (Fig. 4d) revealed the occasional intergranular presence of Zr-Al intermetallic compounds (IMCs), formed either during the synthesis of the $Zr_2AlC$ ceramics [14], or during the LBE exposure as result of Al dissolution in the LBE followed by intergranular LBE penetration into the ZrC grain clusters. The intergranular LBE diffusion into the ZrC grain clusters is supported by the fact that both Bi and Pb were detected along the ZrC GBs, but not inside the ZrC grains (Figs. 4e-4f). Based on the overall findings, there is sufficient evidence of the intergranular penetration of LBE into both $Zr_2AlC$- and ZrC-rich areas in the exposed MAX phase-based ceramics.

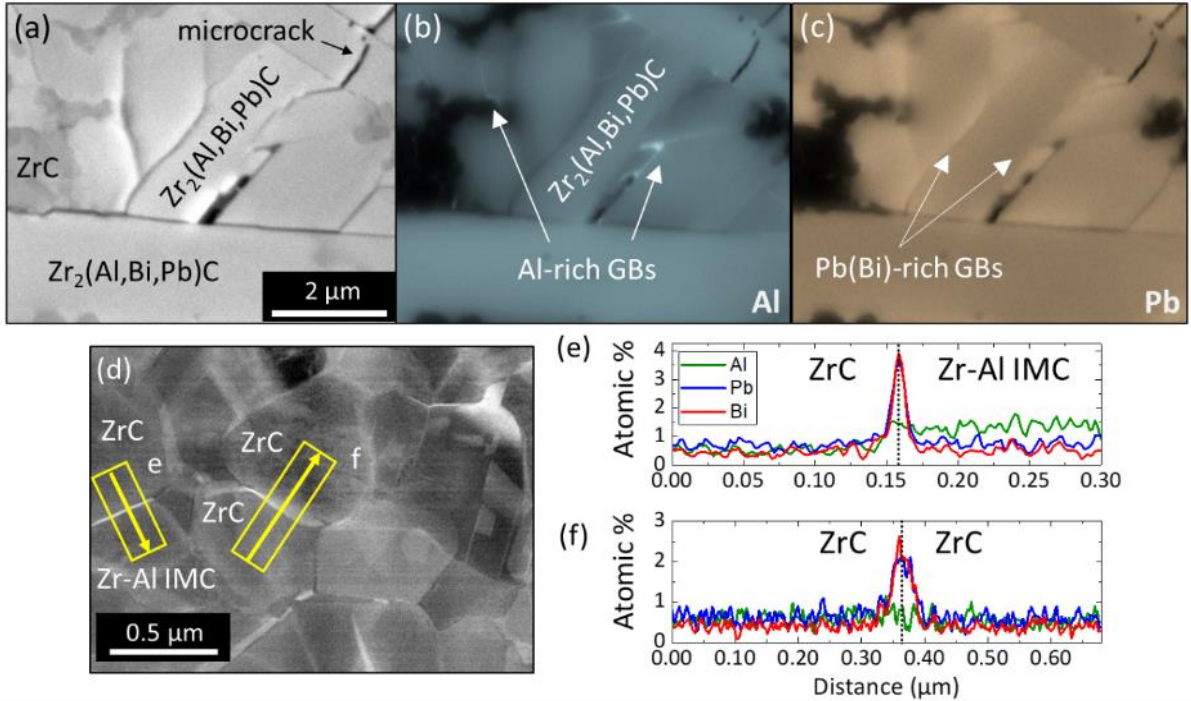

**Fig. 4.** (a) SEM image of $Zr_2(Al,Bi,Pb)C$ and ZrC grains in LBE-exposed $Zr_2AlC$. (b,c) Al and Pb EDS maps of Fig. 4a showing Al and Pb segregation at $Zr_2(Al,Bi,Pb)C$ GBs. (d) STEM image of ZrC grain clusters. (e,f) EDS line scans across two compositionally different ZrC GBs.

To understand the cause of the observed local LBE interaction, cross-sectional FIB trenches were made at different locations on the polished sample facet shown in Fig. 2a. Cross-section *x-x'* is in a seemingly unaffected region and cross-section *y-y'* is in an LBE-affected region (Fig. 2a). In the unaffected region (cross-section *x-x'*, Fig. 5a), ZrC grain aggregates may be seen parallel to the exposed sample surface along with no noticeable LBE infiltration (EDS phase map, Fig. 5b). It is believed that the ZrC grain clusters that happened to be present right under the sample surface acted as a protective layer against LBE ingress, as previously discussed [13]. The clear absence of such ZrC grain clusters in the LBE-affected region (cross-section *y-y'*, Fig. 5c; EDS phase map, Fig. 5d) supports further the postulation that the presence of an LBE-impervious phase, such as ZrC, on the sample surface may prevent or seriously delay the LBE attack. The cross-sections in both affected and unaffected regions showed a thin Zr-Al oxide scale on the sample surface (see supplementary information Fig. S2 for individual elemental maps). This oxide, which has presumably formed in the first stage of the sample immersion in the LBE bath [13], was clearly not protective against LBE attack.

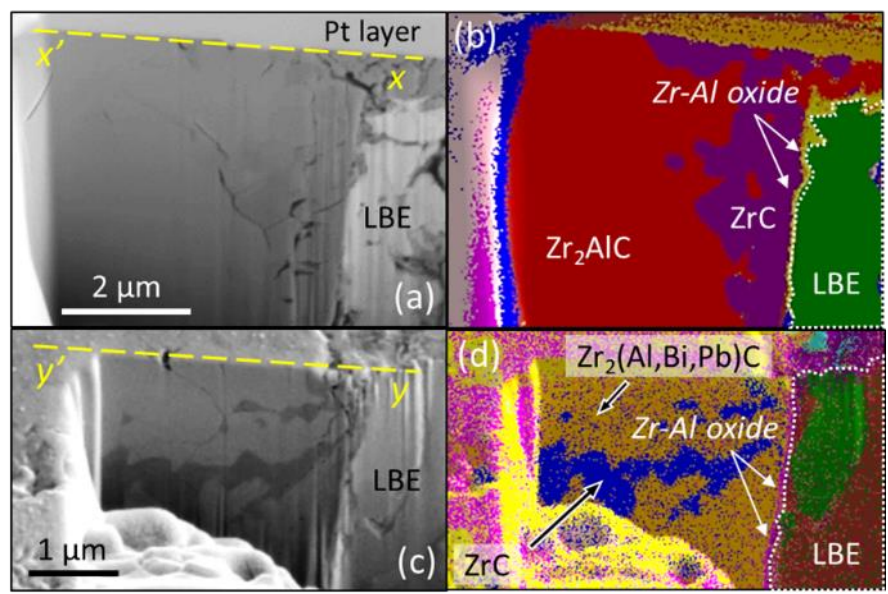

**Fig. 5.** SEM images of FIB cross-sectional trenches on the polished facet of LBE-exposed $Zr_2AlC$: (a) unaffected region (*x-x'* in Fig. 2a) and (c) LBE-affected region (*y-y'* in Fig. 2a). (b,d) EDS phase maps of cross-sections *x-x'* and *y-y'*, respectively.

To investigate the process of LBE infiltration into the $Zr_2AlC$-based ceramic, the 3D chemical distribution of the $Zr_2AlC$/LBE interaction front (marked with three frames in Fig. 2d) was studied in detail by means of APT. Fig. 6 shows the APT elemental distribution profiles of two successive areas/tips inside the transition zone. The diameter of the cylinder from which the profile was computed was 10 nm with 1 nm bin width. The first tip (fully affected area; red-dotted frame in Fig. 2d) contained Bi and Pb, except at its top and bottom left corners (Fig. 6a). A full elemental profile along the *x-x'* arrow is also presented in Fig. 6b, whereas the average composition is provided in Table S2 of the supplementary information as the 'fully affected area'. The composition of the tip from the fully affected area, as obtained by APT (Figs. 6a-6b), corresponds to $\sim Zr_2(Al_{0.71},Bi_{0.17},Pb_{0.12})_{1.14}C_{0.94}$ and is slightly over-stoichiometric in '*A*' elements. The second tip (partially affected area; yellow-dashed frame in Fig. 2d) is divided into two distinctly different halves, where only the top half contains Bi and Pb (Fig. 6c). Moreover, an increased Al content (Al build-up) may be observed at the boundary (i.e., at the Pb/Bi diffusion front) between the two tip halves (Figs. 6c-6d). At the Pb/Bi diffusion front, the Pb and Bi (~7 at% in total) seem to substitute (~7 at%) Al in the Al-overstoichiometric Pb/Bi-free part. This Al overstoichiometry is in agreement with the EDS findings in Figs. 2b and 3b. A third tip, taken from the unaffected area (green-solid frame in Fig. 2d), showed a complete absence of Bi and Pb (results not shown here). APT also revealed the presence of Nb and Hf (Figs. 6a-6d), typical impurities of the ZrH$_2$ powder; these elements are likely to have entered the MAX phase lattice as alloying elements on the *M* (Zr)-site.

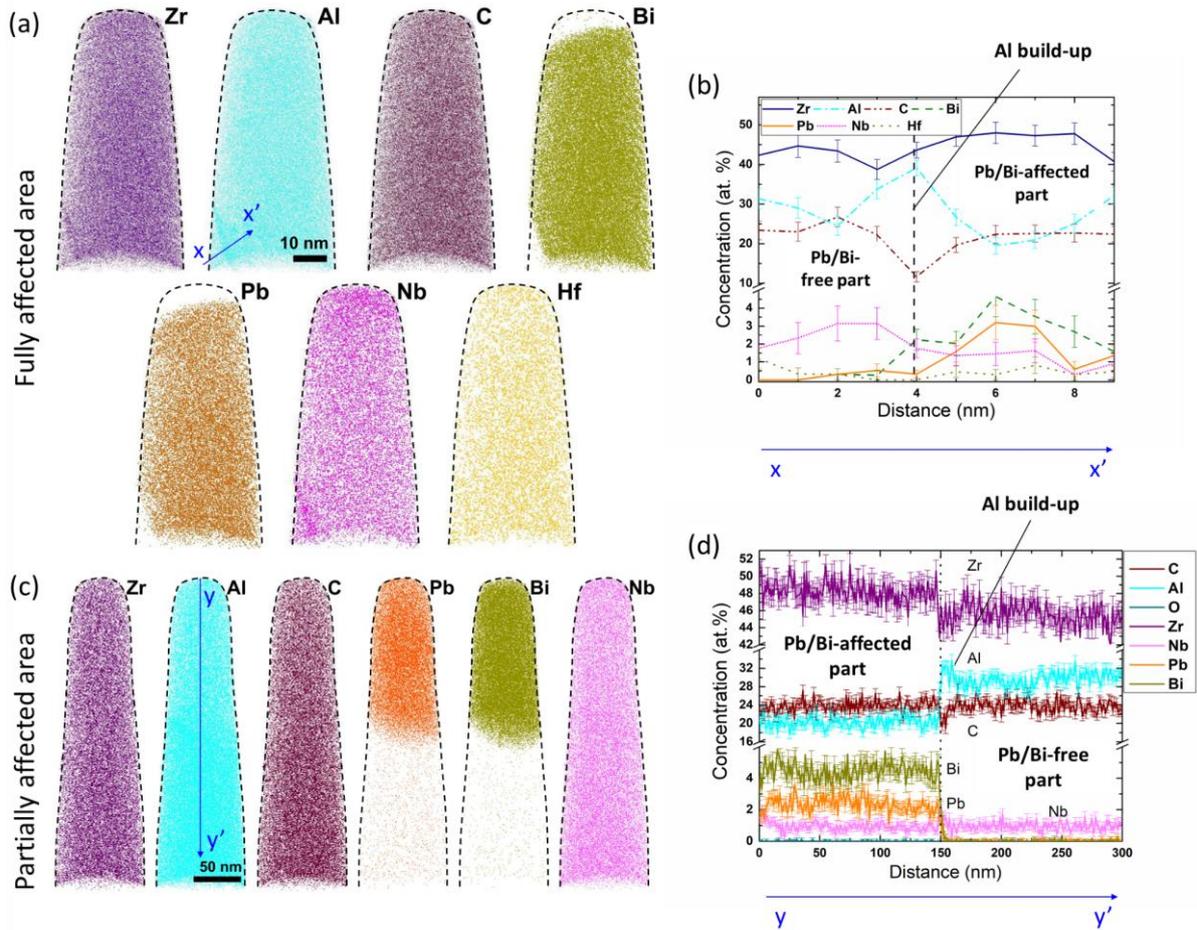

**Fig. 6.** APT elemental distribution profiles acquired from two areas in the transition zone: (a) a fully affected by LBE area (red-dotted frame, Fig. 2d), and (c) a partially affected by LBE area (yellow-dashed frame, Fig. 2d) with their respective elemental profiles (b,d). Arrows *x-x'* and *y-y'* indicate the direction of the two elemental profiles taken across the fully affected region (b) and the partially affected region (d), respectively. Black lines indicate Al build-up at the GBs.

*3.2. Zr$_2$AlC/LBE interaction: TEM/STEM data*

**Fully LBE-affected area**

Since the LBE interaction volume in the exposed ceramic was rather limited, studying the crystal structure of the new MAX phase solid solution required TEM. MAX phase grains in the fully affected area (red-dotted frame in Fig. 2d) contained Bi and Pb, and SAED patterns acquired from these grains revealed a regular MAX phase structure, as shown in Figs. 7c-7d for the $[2\bar{1}\bar{1}0]$ and $[10\bar{1}0]$ zone axes. This was the first indication of Bi and Pb incorporation into the 211 MAX phase structure. HAADF-STEM images of LBE-affected grains (Figs. 7a-7b) clearly show a 211 stacking (Zr$_6$C octahedra separated with Al layers) in the $[2\bar{1}\bar{1}0]$ zone axis, but they also reveal alternating rows of bright atoms at the layers formerly occupied by Al. EDS elemental maps of these regions showed that Al, Bi and Pb preferentially occupy the *A*-site in the 211 MAX phase structure in an ordered manner (Fig. 7e). Bi/Pb-rich *A*-layers alternate with Al-rich *A*-layers, creating the sequential bright and dark contrast in the *A*-layers of the 211 MAX phase structure on the HAADF-STEM images in Figs. 7b and 7e, since the brightness of an atomic column in HAADF-STEM scales with the total atomic number Z of the atoms in that column. The layered order is interrupted by irregularly spaced anti-phase boundaries (red arrows in Fig. 7b) that shift the alternating atomic layer structure by half a unit cell along the *c*-direction. The anti-phase boundaries are not sharp, suggesting that the transition area between two domains is a mixture of these domains. Atomic ordering and anti-phase boundaries were not clearly visible everywhere, probably due to an overlap of neighbouring domains in the $[2\bar{1}\bar{1}0]$ viewing direction or a local absence of the ordering. The lower magnification HAADF-STEM image in Fig. 7a reveals that the observed anti-phase boundaries and atomic ordering are an overall tendency in the 211 MAX phase structure. The compositional differences in the alternating high-Al (low Bi-Pb, *type-1*) and low-Al (higher Bi-Pb, *type-2*) mixed *A*-layers decrease the symmetry of the normal 211 (Zr$_2$AlC) MAX phase. To simulate the effect of such alternation on electron diffraction patterns, a Zr$_2$AlC lattice with *P6$_3$/mmc* space group was converted to P1, discarding symmetry to allow decoupling the two successive *A*-layers. The model with alternating high-Al (*type-1*) and low-Al (*type-2*) layers and the respective simulated SAED patterns can be seen in the supplementary Information Fig. S3. The calculated SAED patterns have no systematic extinctions, and there is good agreement with the FFT (Fast Fourier Transformation) inset image in Fig. 7a representing the small area imaged in HAADF-STEM and the experimental SAED pattern Fig. 7c, both taken along the $[2\bar{1}\bar{1}0]$ zone axis.

The $[10\bar{1}0]$ zone axis (Fig. 7d) shows only the reflections belonging to the original structure, as if no ordering has occurred. This could be due to the occurrence of irregularly spaced nanodomains, overlapping along the viewing direction on a large scale, whereby *type-1* and *type-2* layers become indistinguishable, thus producing the SAED pattern of the $[10\bar{1}0]$ zone axis of the regular Zr$_2$AlC structure. Alternatively, the location, from which this SAED pattern was collected, could have a uniform overall *A*-layer chemistry. Note that in the $[2\bar{1}\bar{1}0]$ zone axis SAED pattern, double diffraction could also have generated the observed extra reflections (e.g., the extra reflection seen between the (0000) and (0002) reflections in Fig. 7c), as compared to the regular Zr$_2$AlC structure.

To roughly estimate the chemical layer composition of the different *A*-layers in Zr$_2$(Al,Bi,Pb)C, a procedure was used similar to that described by Lu *et al.* [32,33]. Details on this procedure are provided in the supplementary information (Figs. S4-S6). Based on the averaged EDS maps of the two chemically different Al/Bi/Pb layers, the darker layer (layer 1) contains approximately Al:Bi:Pb ≈ 66(3):21(2):12(2), and the brighter layer (layer 2) Al:Bi:Pb ≈ 56(3):28(2):16(2). HAADF-STEM image simulations were performed to match the corresponding averaged HAADF-STEM in Fig. 7e to this composition; however, the best match was found for an at% ratio of Al:Bi:Pb 73:18:9 for the darker rows (layer 1), and 50:31:19 for the lighter rows (layer 2). The difference between the values acquired from EDS and HAADF-STEM simulations could be due to the violation of the thin film criterium assumed for the quantification procedure, beam broadening or the channeling effect, all known to affect the outcome for attempts at atomic resolution EDS measurements. The average composition of the new MAX phase solid solution, as determined from overview STEM-EDS maps of this area (Fig. 7e), corresponds to $Zr_2(Al_{0.60}Bi_{0.25}Pb_{0.15})C$. This local STEM-EDS result deviates from the SEM-EDS results of larger areas mentioned earlier. Such variations in local composition can be associated with the distance from the exposed surface and the depth-dependent extent of Bi/Pb ingress.

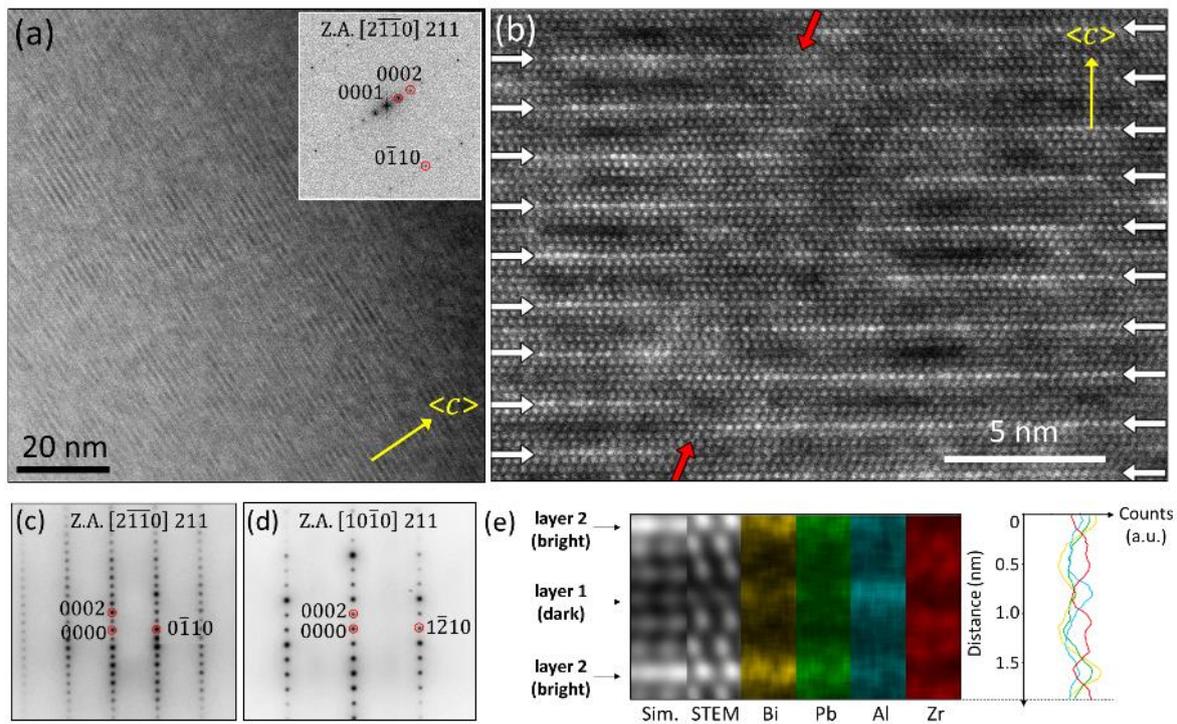

**Fig. 7.** (a) HAADF-STEM overview with corresponding FFT. (b) HAADF-HRSTEM image of the $Zr_2(Al,Bi,Pb)C$ MAX phase solution with alternating bright (white arrows) and dark *A*-layers; an anti-phase boundary is marked by red arrows. SAED patterns showing the (c) $[2\bar{1}\bar{1}0]$ and (d) $[10\bar{1}0]$ zone axes in the 211 MAX phase solid solution. (e) HAADF-STEM simulation performed using QSTEM [19], averaged HAADF-STEM, and elemental EDS maps with their corresponding intensity profiles in counts.

In the fully affected region, the composition of parasitic ZrC grains did not vary, apart from the observed GB decoration by Bi and Pb (Figs. 4d-4f). This implies that ZrC grains are chemically compatible with liquid LBE under the exposure conditions used in this work (500°C, 1000 h, $C_O \leq 2.2\times10^{-10}$ mass%). Twins were commonly observed in ZrC grains (Fig. 8a). Occasionally, planar features characterised by enhanced Bi, Pb and Al contents were observed in the ZrC grains (Figs. 8b-8c); these features were identified as twin boundaries in the ZrC grains. Deformation twin boundaries are considered high stored strain energy interfaces that have also been observed to facilitate LBE ingress into stainless steels tested in similar conditions [3,4]. The presence of Al at ZrC twin boundaries may be attributed to Al that has diffused out of Zr$_2$AlC MAX phase grains, dissolved into the liquid LBE, and diffused along interfaces that facilitate LBE penetration, e.g., grain and twin boundaries.

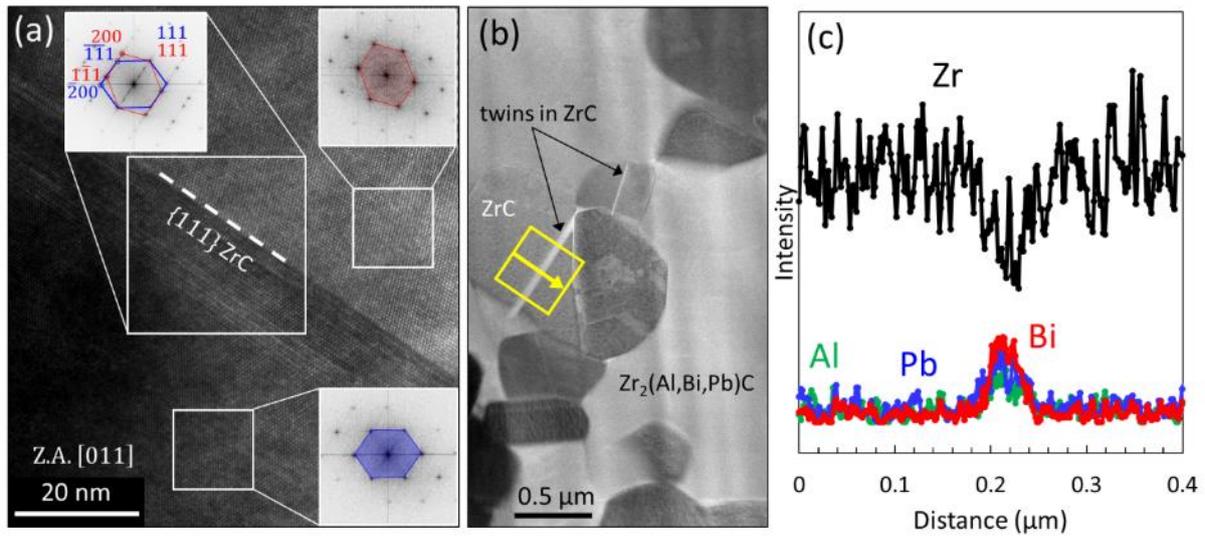

**Fig. 8.** (a) High-resolution TEM (HRTEM) image of the commonly observed {111} twins in ZrC grains with FFT inserts. Z.A. stands for 'zone axis'. (b) STEM image of ZrC grains found in the fully affected area (red-dotted frame, Fig. 2d) with Bi/Pb/Al-containing twin boundaries, as confirmed by the EDS line scan across one of the twin boundaries (c).

**Partially LBE-affected area**

The intriguing feature of this area are the Bi/Pb-containing, bright 'stripes' that appear inside otherwise unaffected Zr$_2$AlC grains (Figs. 9a and 9c). Similar features are visible in the SEM image of Fig. 2d and have been previously reported [13]. The clear difference in compositional contrast indicates the presence of Bi and Pb in the LBE-affected MAX phase grains compared to the unaffected Zr$_2$AlC grains. Pb/Bi atoms are either present in the whole grain or only locally, forming bright 'stripes' within the grain, that are believed to have resulted from the preferential Pb/Bi diffusion along stacking faults (SFs); in fact, the formation of numerous SFs is typical for the powder metallurgical synthesis of MAX phase-based ceramics. As an example, weak beam TEM images of SFs and dislocations in unaffected grains of the ceramic bulk are given in the supplementary information (Fig. S7). The Bi/Pb-affected areas are visible in Figs. 9a and 9c as bright slabs with boundaries parallel to the (0001) planes of the Zr$_2$AlC MAX phase (orientation verified using SAED patterns, not shown here). The preferential onset of Bi/Pb penetration parallel to the basal planes may be attributed to the fact that basal planes are densely packed planes kept together by weaker *M-A* bonds [7]. Therefore, the basal planes serve as fast diffusion tracks for heavy atoms, such as Pb and Bi. The LBE penetration along basal planes and SFs is followed by the Pb/Bi diffusion towards other planes, further affecting the grain. Fig. 9a captures an early step in the LBE ingress process, where the heavy liquid metal atoms follow SFs (bright stripes) in the Zr$_2$AlC grain. Interestingly, the EDS point analysis of this LBE-affected 'stripe' determined a Zr$_2$(Al$_{0.93}$,Bi$_{0.07}$)C$_y$ tentative composition without any Pb, the absence of which is also visible in the EDS line scan of Fig. 9b. This may be attributed to the smaller atomic diameter of Bi, which possibly initiates the LBE ingress, opening up the lattice to accommodate the larger Pb atoms. At this early stage, the reduction in the Al content of the $Zr_2AlC$ MAX phase is not yet visible (Fig. 9b). The non-attacked (dark-colored) parts of the grains contained no Pb or Bi. A more advanced step in the process of LBE ingress is shown in Fig. 9c, where the loss of Al is visible, and Pb is also present in the LBE-affected 'stripe' (Fig. 9d). Overall, STEM-EDS results (Fig. 9) acquired from this transition zone clearly showed that Bi and Pb ingress into the MAX phase grains started preferentially along the basal planes with occasional protrusions along other planes as the LBE attack progresses (Fig. 9d).

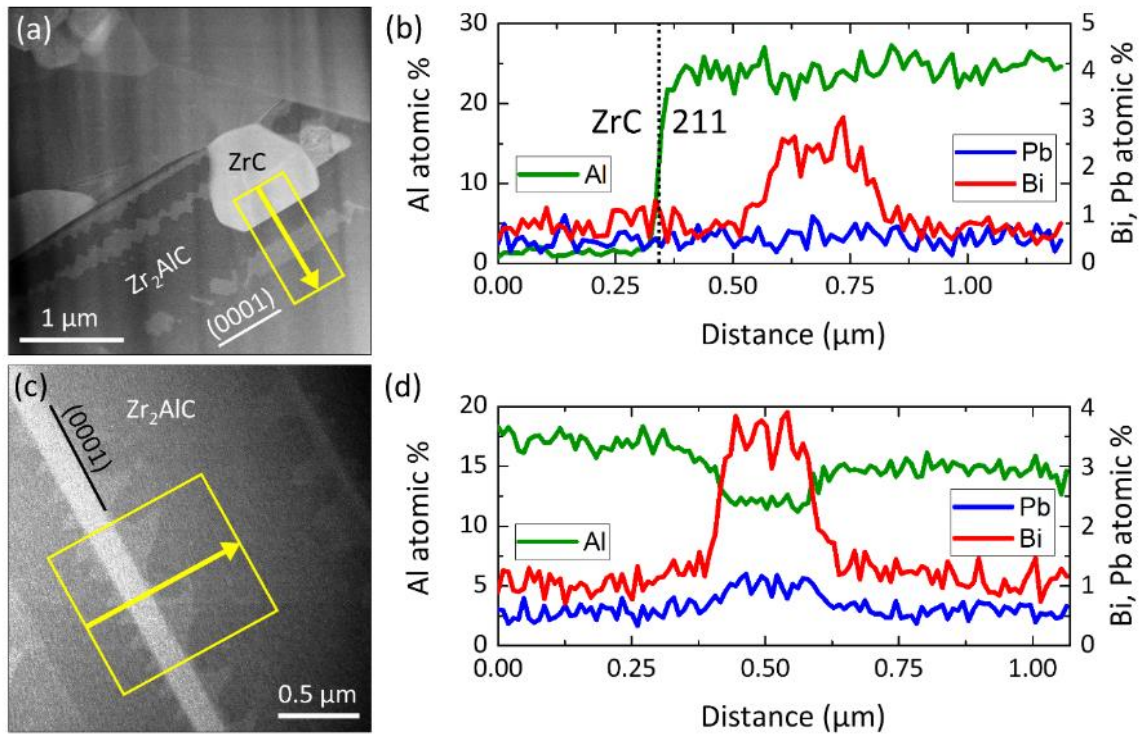

**Fig. 9.** (a,c) STEM images of the progress of LBE ingress at the $Zr_2AlC$/LBE interaction front. EDS line scans showing (b) the early Bi penetration, and (d) the penetration of both Bi and Pb into the $Zr_2AlC$ basal planes and the ensuing decrease in the Al content.

**Unaffected area**

The unaffected area (green-solid frame in Fig. 2d) contained MAX phase grains without Bi and Pb. Contrary to what was expected, these grains had substantially different SAED patterns than the $Zr_2AlC$ and solid solution MAX phase grains in the other two areas (fully or partially LBE-affected regions). Tilt series of SAED patterns were collected, each of them having extra dim reflections in addition to the bright regular MAX phase reflections, as shown in Fig. 10. HRSTEM imaging performed along the $[2\bar{1}\bar{1}0]$ and $[0001]$ zone axes is presented in Figs. 10a, 10c and 10e. The sample was beam sensitive, especially when imaged along the $[0001]$ zone axis, causing superlattice reflections (Figs. 10i and 10j) to quickly disappear and also be absent on the FFTs after subsequent HRSTEM imaging or EDS mapping. However, from the high resolution images along these two zone axes, it was possible to deduce some key information about this Pb/Bi-free, somehow modified, Zr$_2$AlC MAX phase structure.

In Figs. 10a and 10c, both HAADF and ABF STEM images are given along the $[2\overline{1}\overline{1}0]$ zone axis. ABF imaging allows visualizing lighter elements such as C, whereas these are out of reach for HAADF-STEM imaging. The FFT of the HAADF-STEM image (Fig. 10f) did not show any extra reflections, whereas the FFT of the ABF image (Fig. 10h) showed faint extra reflections at $\frac{1}{2}c^*$, as marked by the white arrows. In Figs. 10b and 10d, filtered HAADF and ABF-STEM images are given, with magnified sections included in Fig. 10g. The filtered ABF-STEM image indeed showed that atomic rows inside the Zr$_6$C layers (marked by white arrows in Figs. 10d and 10g) were missing, whereas neighbouring rows of Zr$_6$C (indicated by yellow arrows in Fig. 10d and 10g) showed the presence of atoms on the C sites, as normally expected. Due to the electron beam sensitivity, no EDS measurements were successful in determining whether that atom was C or anti-site Al. The fact that these extra features were not visible in HAADF-STEM images indicates a low total Z of the 'defect atoms', which also allows both light atoms (C) or very low amounts of Al. When examining the HRSTEM images along the $[0001]$ zone axis (Fig. 10e), bright columns of atoms were observed at the location where C atoms should occupy the MAX lattice, as indicated by red arrows. C atoms are expected to be invisible in HAADF-STEM imaging due to the approximate $Z^2$ (Z = atomic number) contrast. Appearance of bright atomic columns at the C sites (in the $[0001]$ zone axis) suggests that, apart from the possible vacancies observed in ABF imaging in the $[2\overline{1}\overline{1}0]$ zone axis, there might be anti-site Al defects on the C sites. Due to the beam sensitivity, observation of periodic repetitions of these potential defects (that can generate superlattice reflections) or elemental identification via EDS was not possible. As a result of the inconsistent Al concentrations measured in grains without Bi/Pb (ranging from Zr$_2$AlC$_y$ to Zr$_2$Al$_{1.33}$C$_y$), and also in grains of thermally annealed Zr$_2$AlC (ranging from Zr$_2$Al$_{0.48}$C$_y$ to Zr$_2$Al$_{1.44\pm0.18}$C$_y$) (as discussed below), it seems that both an under- and over-stoichiometry in Al is possible in Zr$_2$AlC.

In order to assess whether the observed superlattice was related to the neighbouring LBE-affected zone, additional FIB foils of MAX phase grains from the centre of the bulk sample were investigated. Here, the grain compositions were close to the original Zr$_2$AlC$_y$. Again, the same superlattice was observed where the orientation of the unaffected grains enabled to collect tilt series covering the full range between the $[2\overline{1}\overline{1}0]$ and the $[0001]$ zone axes, so as to identify the modulation vectors. The superlattice reflections could then be indexed by the modulation vectors $q_1$, and $q_2$:

$$q_1 = -\frac{1}{6}a^* + \frac{1}{3}b^* \qquad (1)$$

$$q_2 = \frac{1}{6}a^* + \frac{1}{3}b^* \qquad (2)$$

Although the modulation vectors allow indexing all the extra spots observed experimentally, no model for the crystal structure could be identified to perfectly match the experimentally observed intensities. At the moment, it is impossible to provide concrete evidence of the effect of Al (under- and over-stoichiometry) on the observed superlattice. Considering the open literature, the superlattice could have arisen due to C vacancy ordering. C vacancy ordering is common in substoichiometric ZrC (as in ZrC$_{0.61}$) [34], and was also reported in Nb$_{12}$Al$_3$C$_8$ [35] and V$_{12}$Al$_3$C$_8$ [36] MAX phases. More work is needed to investigate the intrinsic thermal stability of the Zr$_2$AlC MAX phase, as the ordering of the (probably temperature-dependent) population of C vacancies due to prolonged annealing might affect the material's radiation swelling behaviour in a nuclear service environment.

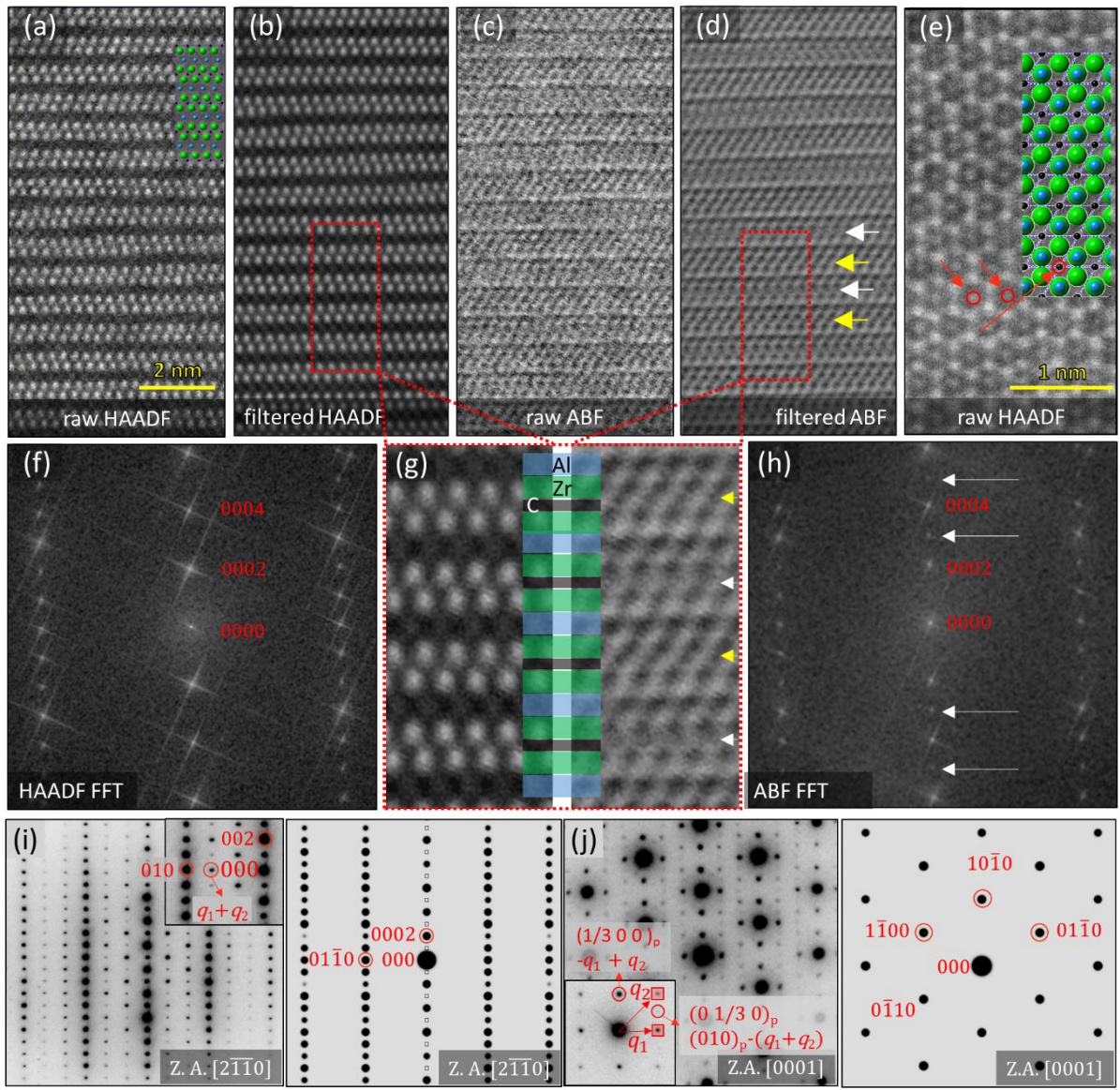

**Fig. 10.** LBE-unaffected Zr$_2$AlC grains. Imaging along the $[2\bar{1}\bar{1}0]$ Z.A., (a) raw STEM-HAADF image, (b) filtered HAADF image, (c) raw ABF image, (d) filtered ABF image, (e) HAADF-STEM image in the $[0001]$ Z.A. with Zr$_2$AlC unit cells added as a guide. FFTs of (f) HAADF and (h) ABF-STEM images along the $[2\bar{1}\bar{1}0]$ Z.A. (g) Magnified inserts from filtered HAADF and ABF images. Experimental and simulated SAED patterns of Zr$_2$AlC at the (i) $[2\bar{1}\bar{1}0]$ and (j) $[0001]$ Z.A.

In this part of the work, structural and chemical changes that were not necessarily associated with the LBE ingress (as no Pb or Bi was detected in the corresponding regions) were observed in the Zr$_2$AlC MAX phase. This suggests that this material is intrinsically unstable at the experimental LBE exposure conditions. At the LBE exposure conditions (500°C, 1000 h), the Zr$_2$AlC phase transforms to a supercell structure. To study these intrinsic material changes, a batch of the Zr$_2$AlC-based ceramics was vacuum-annealed at 500°C for 1000 h, as discussed in the following section.

*3.3. Vacuum-annealed Zr$_2$AlC*

Zr$_2$AlC ceramics were vacuum-annealed in sealed quartz tubes at 500°C for 1000 h. The XRD spectra before and after annealing are compared in Fig. S8 of the supplementary information. No significant changes in the XRD patterns were observed, but the weight percentages before and after annealing changed slightly with respect to the Zr$_2$AlC amount (~10 wt% reduction in Zr$_2$AlC with an equivalent increase in ZrC and Al$_3$Zr$_2$). No significant lattice parameter changes were measured by Rietveld refinement, while SEM examination of the as-annealed sample surface showed basal plane delaminations in the MAX phase grains and the presence of dark Al$_2$O$_3$ grains (Fig. 11a). The basal plane delaminations reflect the loss of Al due to the vacuum environment, while the formation of Al$_2$O$_3$ indicates that the Zr$_2$AlC ceramic acted as a getter for the oxygen impurities in the quartz tube used for sample encapsulation. MAX phase decomposition according to the general reaction *[M$_{n+1}$AX$_n$ → M$_{n+1}$X$_n$ + A]* has been reported for Ti$_2$AlC, Ti$_3$AlC$_2$, Ti$_3$SiC$_2$, Ti$_2$AlN, Ti$_4$AlN$_3$ [37], Ta$_4$AlC$_3$ [38] (via Al sublimation), Zr$_2$(Al$_{0.42}$,Bi$_{0.58}$)C (upon pressure application, [39]), and Cr$_2$AlC (in low-oxygen content Ar, [40]). In all cases, however, the reported decomposition temperatures were above 1200°C [37–40], while, in this work, the Zr$_2$AlC ceramic was vacuum-annealed at 500°C, which is too low to trigger a conventional MAX phase decomposition. The observed (limited) transformation of Zr$_2$AlC to ZrC and Al$_3$Zr$_2$ could only be attributed to elemental redistribution occurring during vacuum annealing, which presumably destabilised the MAX phase crystal structure, favoring the formation of the competing phases ZrC and Al$_3$Zr$_2$. It is well understood that the inherent propensity of the Zr$_2$AlC MAX phase towards its decomposition into the competing phases ZrC and Al$_3$Zr$_2$ can be associated with the high trigonal prism distortion value (i.e., 1.101 [41,42]) in the Zr$_2$AlC crystal lattice. The elemental redistribution observed in this work due to vacuum annealing (i.e., superlattice formation associated with the redistribution of C and Al in the bulk, Al loss and O ingress close to the sample surface) might have destabilised further the MAX phase crystal lattice, causing a partial Zr$_2$AlC → ZrC + Al$_3$Zr$_2$ phase transformation.

EDS point analysis of the as-annealed sample surface indicated the Zr$_2$AlC MAX phase to be substoichiometric in Al, with an approximate composition of Zr$_2$Al$_{0.48}$C$_y$, and also revealed the formation of isolated Al$_2$O$_3$ grains (Fig. 11a). As already mentioned, the possibility of Al loss close to the sample surface cannot be excluded in a vacuum environment, while the local formation of Al$_2$O$_3$ grains could be attributed to oxygen impurities in the quartz tube. On the other hand, EDS measurements on metallographic cross-sections (Fig. 11b) showed a Zr$_2$Al$_{1.4\pm0.2}$C$_y$ composition, i.e., Al-overstoichiometric, and similar to the bulk composition of LBE-exposed Zr$_2$AlC (Figs. 2b, 3b and 6d).

The TEM FIB foil from the as-annealed surface also showed the similar superlattice reflections as observed in the bulk of the LBE-exposed sample and in the Pb/Bi-free grains in the transition zone and were indexed using the same modulation vectors $q_1$ and $q_2$ (Fig. 11d). Since these superlattice reflections were not visible in the XRD patterns of both LBE-exposed or annealed Zr$_2$AlC samples, their crystal structure could not be refined. SFs were commonly observed in large MAX phase grains (Fig. 11c) and superlattice reflections were visible in SAEDs, as shown in Fig. 11d for the $[2\bar{1}\bar{1}1]$ zone axis, but they were also seen in other investigated zone axes.

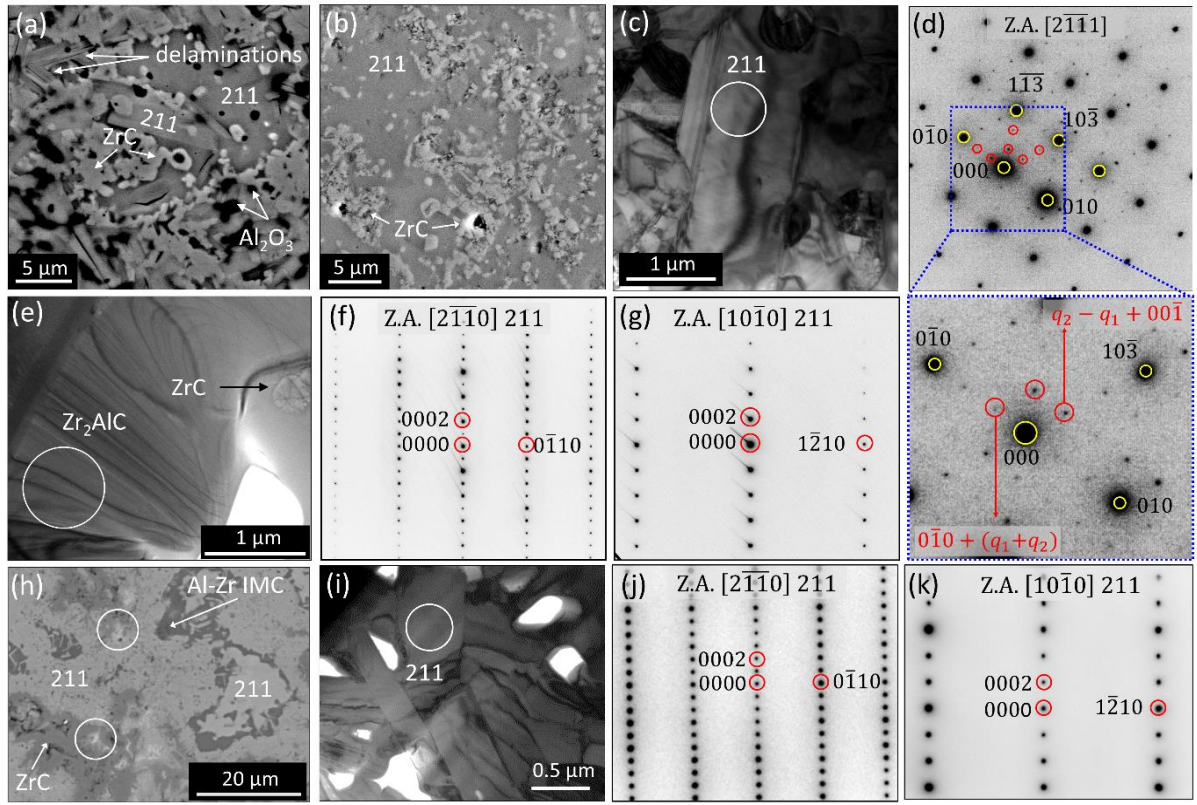

**Fig. 11.** SEM images of vacuum-annealed $Zr_2AlC$: (a) top surface, (b) cross-section. (c) BF-TEM and (d) SAED pattern from the circle in Fig. 11c: $[2\bar{1}\bar{1}1]$ Z.A. of $Zr_2AlC$ (the magnified part of the SAED pattern shows superlattice reflections). Pristine $Zr_2AlC$: (e) BF-TEM image and (f,g) SAED patterns at the $[2\bar{1}\bar{1}0]$ and $[10\bar{1}0]$ Z.A. Hot-pressed $Zr_2(Al,Bi,Pb)C$: (h) SEM image (circles around the Bi/Pb-rich IMC), (i) BF-TEM image and (j,k) SAED patterns at the $[2\bar{1}\bar{1}0]$ and $[10\bar{1}0]$ Z.A.

Noticing the possible inherent instability of the $Zr_2AlC$ MAX phase, a piece (stored at RT for 3 years) from the original batch of as-sintered $Zr_2AlC$-based ceramic was also investigated (Figs. 11e-11g). No superlattice reflections were observed and the extracted thin foil, which was substoichiometric in Al with an approximate composition of $Zr_2Al_{0.67}C_y$. This composition was different than that measured and initially reported for the as-synthesised $Zr_2AlC$ ceramics [14], where the Zr:Al:C atomic ratio in the $Zr_2AlC$ MAX phase was determined by XPS to be 52.1:23.1:24.8. That composition was in agreement with stoichiometric $Zr_2AlC$, for which the same (nominal) ratio is 50:25:25. The observed overall changes that occurred upon RT storage in both as-sintered and annealed/LBE-exposed at 500°C samples, suggest a low-temperature instability of the $Zr_2AlC$ MAX phase.

## 3.4. DFT calculations on Zr$_2$(Al,Bi,Pb)C

The calculated substitutional energy, $\Delta E_{s,Al}^{Bi+Pb}$, upon partial substitution of Al for Bi and Pb at 0 K, is shown in Fig. 12a for the Zr$_2$(Al$_{1-x}$,Bi$_{2x/3}$,Pb$_{x/3}$)C general stoichiometry. $\Delta E_{s,Al}^{Bi+Pb}$ is negative for $x < 1$, indicating that the substitution of Al for Bi and Pb is energetically favorable. A minimum in $\Delta E_{s,Al}^{Bi+Pb}$ at 0 K is found around $x = 0.375$ (62.5 at% Al). This corresponds well with the range of compositions measured by EDS and APT, where the Al/A-element ratio varied between 0.50 and 0.79. However, the minimum in $\Delta E_{s,Al}^{Bi+Pb}$ is shifted to higher $x$ values with increasing temperature, when the entropy contribution is taken into account. At 1273 K, the minimum of $\Delta E_s^{Bi.Pb}[T]$ is close to $x = 0.5$, corresponding to Al:Bi:Pb $\approx$ 3:2:1. This also indicates that at the LBE exposure temperature of 500°C, both solubility and stability of Bi and Pb in the Zr$_2$AlC MAX lattice are higher than at RT, which might lead to a supersaturation of these elements at RT. Fig. 12a shows the results for $\Delta E_s^{Bi.Pb}$ at 0 K (blue line) and with entropy contribution at 1273 K (red line), along with the Al, Bi, and Pb contents indicated as as $x$, $y$ and $z$, respectively, in the Zr$_2$(Al$_x$,Bi$_y$,Pb$_z$)C stoichiometry (right axis).

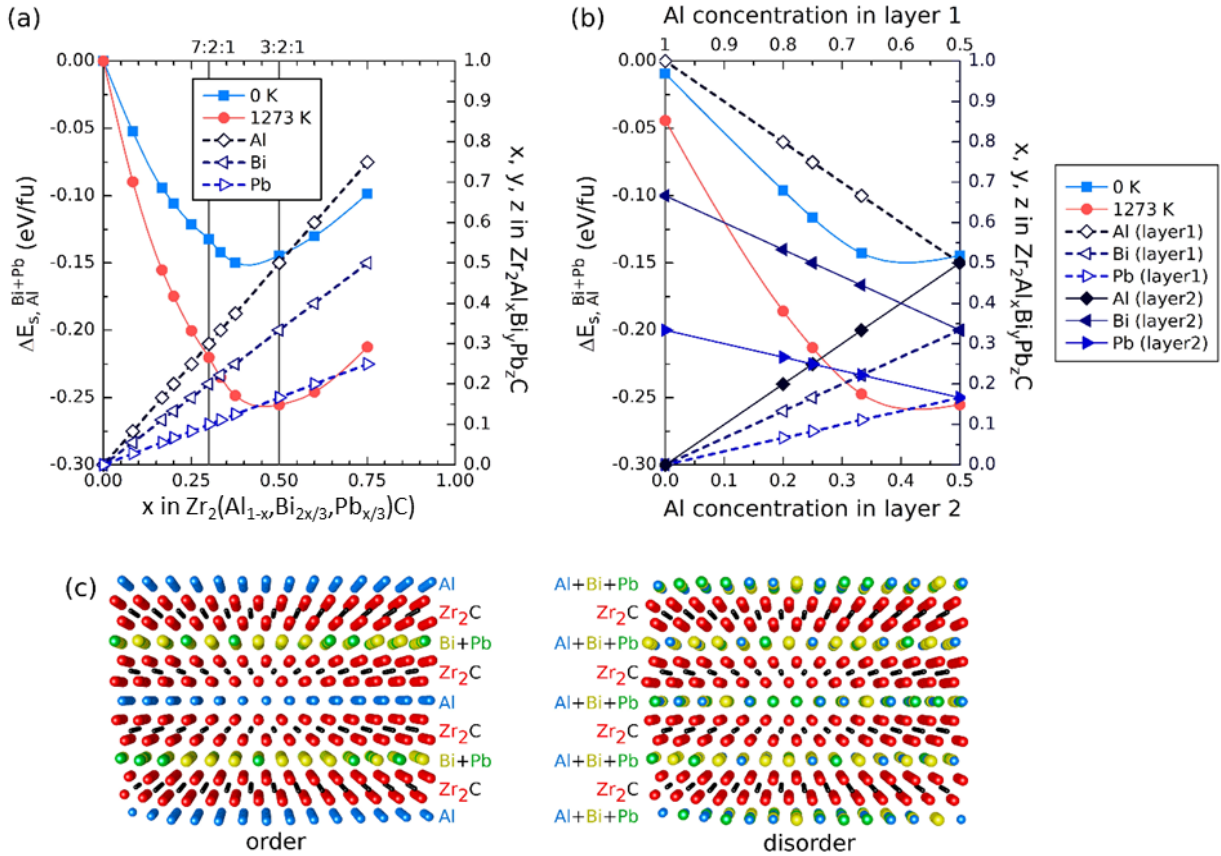

**Fig. 12.** Substitutional energy, $\Delta E_{s,Al}^{Bi+Pb}$, upon substituting Al for Bi and Pb in Zr$_2$AlC: (a) in a Zr$_2$(Al$_{1-x}$,Bi$_{2x/3}$,Pb$_{x/3}$)C general stoichiometry with varying $x$ at 0 K (blue line), and with entropic contribution up to 1273 K (red line), and (b) in different A-layers (layers 1 and 2), with varying Al concentration in each layer (at a fixed overall Bi:Pb $\approx$ 2:1) at 0 K (light blue line with square symbols), and with entropic contribution at 1273 K (red line), along with the concentration of Al, Bi and Pb on the different *A*-sites. (c) Ordered (pure Al and mixed Bi/Pb layers) and disordered models (intermixed layers) used for the calculations shown in Fig. 12b.

Structurally, the partial substitution of Al for Bi and Pb in Zr$_2$AlC, up to *x* = 0.5, results in an almost linear increase of lattice parameters *a* and *c*, as shown in Fig. S9 of the supplementary information. At *x* = 0.5, *a* and *c* have increased by 1.2% and 0.8%, respectively, when compared to Zr$_2$AlC. Further substitution of Al for Bi and Pb results in a larger increase in *a*, whereas *c* starts to decrease. This can be partially associated with the smaller atomic radius of Al as compared to the radii of Bi and Pb that could be accommodated in the *A*-layer. Since the Zr-*A* layer is not as close-packed compared to the Zr-C layer, this makes the *A*-layer rather flexible for elemental substitution. For *x* > 0.5, where the majority of atoms in the *A*-layer consists of the larger Bi and Pb atoms, the structure expands in-plane to minimize size-related strains. In addition, the bond length distribution in the disordered solid solution has been extracted up to a cut-off of 3.8 Å and is presented in Fig. S10 of the supplementary information.

Up to this point, the main focus was to investigate the substitution of Al by Bi and Pb. For investigating the compositional differences observed in STEM in the alternating *A*-layers of the Zr$_2$(Al,Bi,Pb)C solid solution, an overall Al:Bi:Pb ratio of 3:2:1 was assumed. This ratio was observed in TEM foils similar to those shown in Figs. 2d and 3d. However, the concentration of the individual *A*-layers has varied. The calculated substitutional energy, $\Delta E_{s,Al}^{Bi+Pb}$, and specified concentrations of Al, Bi, and Pb within the alternating layers are shown in Fig. 12b. The schematics in Fig. 12c show two of the structures used for the calculations in Fig. 12b, where the first model (indicated as 'ordered') represents the case of alternating layers of pure 'Al' and 'Bi+Pb', while the second model (indicated as 'disordered') represents alternating intermixed *A*-layers containing Al, Bi and Pb in a 3:2:1 ratio.

The calculated substitutional energy, $\Delta E_{s,Al}^{Bi+Pb}$, in Fig. 12b suggests that the ordered case of alternating layers of pure Al with mixed Bi/Pb layers is less favored as compared to the disordered case with a distribution of Al, Bi, and Pb in every *A*-layer. This means that $\Delta E_{s,Al}^{Bi+Pb}$ decreases when Al is mixed into the Bi/Pb-rich layers. A minimum $\Delta E_{s,Al}^{Bi+Pb}$ is reached at a concentration of 33 at% Al in layer 2 (the Bi/Pb-rich layer that appears bright in the HRSTEM images), and when layers 2 and 1 have Al:Bi:Pb ratios of 3:4:2 and 6:2:1, respectively. Although the fully ordered model is not favorable, this tendency towards alternating mixed *A*-layers with varying Al:Bi:Pb ratios (i.e., higher Al/lower Bi-Pb layers followed by lower Al/higher Bi-Pb layers) confirms the experimental findings acquired by HRSTEM and EDS. With increasing temperature, *T* > 0 K, the disordered case becomes more favorable due to a larger entropic contribution.

*3.5. Reactive hot-pressed Zr$_2$(Al,Bi,Pb)C*

The Zr$_2$(Al,Bi,Pb)C MAX phase solid solution that formed in-situ during the LBE exposure was also synthesized by conventional reactive hot pressing at 1250°C for 30 min under 30 MPa. An Al:LBE molar ratio of 1:1 was used in the starting powder mixture. The SEM investigation of a metallographic cross-section of this ceramic revealed a relatively fine-grained microstructure consisting of brighter Zr$_2$(Al,Bi,Pb)C MAX phase grains as well as darker ZrC, Al-Zr IMC and occasional Bi/Pb-rich IMCs grains (Fig. 11h). The composition of the MAX phase grains, as determined by EDS point analysis, was approximately $Zr_2(Al_{0.69},Bi_{0.19},Pb_{0.12})_{1.08}C_y$, i.e., similar to the composition of LBE-affected 211 MAX phase grains, as determined by TEM, SEM-EDS and APT. The detection of *A*-overstoichiometric 211 grains as well as residual IMC and ZrC grains indicates that the synthesis reaction was incomplete and could be further optimized.

TEM examination of the $Zr_2(Al,Bi,Pb)C$ ceramic revealed regular MAX phase SAED patterns, as shown in Figs. 11j and 11k. No extra reflections were detected in these SAED patterns. As with the TEM findings from LBE-affected areas, the missing *hkil: l=2n+1* reflections in the $[10\bar{1}0]$ zone axis indicate either a uniform *A*-layer chemistry (which is probable considering that the sintering temperature of 1250°C was higher than the LBE exposure temperature of 500°C, where $Zr_2(Al,Bi,Pb)C$ formed in-situ) or an overlap of many nanoscale domains with different *A*-layer compositions. The finer grain size observed in this sample could probably be associated with the low sintering temperature (1250°C). The Bi/Pb-rich IMC phase detected by SEM-EDS and marked by circles in Fig. 11h proved more sensitive to the Ga ion beam during FIB milling, and its preferential removal left holes in the final foil.

As mentioned earlier, the bulk synthesis of the $Zr_2(Al,Bi,Pb)C$ MAX phase solid solution enabled refinement of its crystal structure using XRD and NPD techniques, something that was not possible in the as-exposed ceramic due to limited volume of the in-situ formed solid solution.

*3.6. Neutron diffraction of reactive hot-pressed $Zr_2(Al,Bi,Pb)C$*

The RT XRD pattern of hot pressed $Zr_2(Al,Bi,Pb)C$ (Fig. S11a of the supplementary information) consisted of discrete reflections that could be indexed by the hexagonal structure (space group *P6$_3$/mmc*) with $a \approx 3.35$ Å and $c \approx 14.54$ Å lattice parameters. These reflections corresponded to the major phase, which was the 211 ($Zr_2(LBE_{0.5},Al_{0.5})C$) MAX phase solid solution. Another set of reflections that belonged to a competing phase could be indexed by a cubic structure (space group *Fm-3m*) with lattice parameter $a \approx 4.68$ Å, which corresponds to ZrC [43]. Two more minor phases were identified as $Zr_2Al_3$ [44], and $ZrAl_3$ [45]. The RT NPD pattern (Fig. S11b) showed reflections that could be indexed in a similar way as those observed by XRD, with the exception of the $ZrAl_3$ minor phase, for which no reflection was distinctly present. The joint Rietveld analysis identified four phases (Fig. S11): the two major phases were $Zr_2(LBE_{x\approx0.5},Al_{y\approx0.5})C$ and ZrC, and the two minor phases were $Zr_2Al_3$ and $ZrAl_3$ IMCs.

The starting $Zr_2(LBE_{x\approx0.5},Al_{y\approx0.5})C$ structure was assumed to have a regular MAX phase crystal structure in the hexagonal *P6$_3$/mmc* space group with the following atomic positions: Zr in *4f*(1/3,2/3,z≈0.58), LBE/Al in *2c*(1/3,2/3,1/4), and C in *2a*(0,0,0). The starting ZrC structure was assumed to have the cubic *Fm-3m* structure with Zr in *4a*(0,0,0) and C in *4b*(1/2,1/2,1/2). The two minor impurity phases were assumed to have the structures reported in literature [44,45]. The refined parameters of the structural model are summarized in Table 1.

The proposed model generated calculated XRD and NPD patterns that were in very good agreement with the observed RT data (Fig. S11). The model fit to the XRD [NPD] data agreed with the presence of 72(2) [73(4)] wt% of $Zr_2(LBE_{x\approx0.5},Al_{y\approx0.5})C$ MAX phase, 21(1) [23(2)] wt% ZrC, 5(1) [4(2)] wt% $Zr_2Al_3$, and 2(1) [0] wt% $ZrAl_3$. The excellent agreement of the phase content, as derived by an unconstrained scale factor refinement for each pattern, strongly indicates the validity and robustness of the suggested model.

**Table 1.** Joint Rietveld analysis of experimental XRD and NPD patterns. Refined structural parameters are: the unit cell parameters ($a$ and $c$), weight percent of phases (wt%), isotropic atomic thermal displacement parameter ($B_{iso}$) and individual phase ($R_F$), whole pattern expected ($R_{exp}$), and whole pattern weighted ($R_{wp}$) agreement factors [25].

| Phase | Parameter | | Refined values |
|---|---|---|---|
| Zr$_2$(LBE$_x$,Al$_y$)C$_\delta$ | $a$ (Å) | | 3.3472(2) |
| | $c$ (Å) | | 14.5401(8) |
| | wt.% | NPD | 73(4) |
| | wt.% | XRD | 72(1) |
| | Zr | z | 0.5834(5) |
| | | $B_{iso}$ (Å$^2$) | 1.25(6) |
| | (LBE$_x$,Al$_y$) | x | 0.59(2) |
| | | y | 0.5(2) |
| | | $B_{iso}$ (Å$^2$) | 7.6(2) |
| | C | δ | 0.81(4) |
| | | $B_{iso}$ (Å$^2$) | 0.6(2) |
| | $R_F$ | NPD | 13.6 |
| | | XRD | 4.94 |
| ZrC$_\beta$ | $a$ (Å) | | 4.6764(3) |
| | wt% | NPD | 23(2) |
| | | XRD | 21(1) |
| | Zr | $B_{iso}$ (Å$^2$) | 1.8(2) |
| | C | β | 0.63(4) |
| | | $B_{iso}$ (Å$^2$) | 0.9(4) |
| | $R_F$ | NPD | 7.05 |
| | | XRD | 1.37 |
| $R_{exp}$ | | NPD | 6.89 |
| | | XRD | 3.38 |
| $R_{wp}$ | | NPD | 8.03 |
| | | XRD | 10.9 |

The occupancies of the different elements in the Zr$_2$(LBE$_{x\approx0.5}$,Al$_{y\approx0.5}$)C MAX phase solid solution were refined using several scenarios, four of which should be discussed. First, Pb and Bi were considered to exist on both the Zr as well as the Al sites. The refinement under this assumption resulted in a full Zr occupancy and zero LBE occupancy on the *M* metal (*4f*) site, and significant occupancy on the *A* metal (*2c*) site. Second, Pb, Bi, and Al occupancies were refined on the *A* metal site, resulting in divergence (as expected). Third, the Al occupancy was fixed at 0.5 and the unconstrained Pb and Bi occupancies were refined. This resulted in negligible Bi occupancy and ~0.5 site occupancy for Pb. However, the uncertainties on those values were ~0.5 as well, rendering this result unlikely. In disagreement with this result, other methods such as APT and EDS have shown the presence of Bi in the MAX phase. Finally, the reported model in Table 1, the one showing the best agreement with the observed data, was achieved by fixing the Bi/Pb ratio to 1, and refining the unconstrained LBE and Al occupancies. In all models, the C occupancy was individually refined, resulting in vacancies in both the MAX phase and ZrC.

In summary, the final formula of the MAX phase was refined to $Zr_2(LBE_{0.59(2)},Al_{0.5(2)})C_{0.81(4)}$, with the following lattice parameters: $a$ = 3.3472(2) Å and $c$ = 14.5401(8) Å. Compared to $Zr_2AlC$, there is an expansion along the $a$ and contraction along the $c$ direction. The compositional differences between this MAX phase solid solution and the one observed in LBE-affected areas could be linked with differences in their synthesis temperature (e.g., the solid solution formed at 1250°C in the hot pressed material and at 500°C in the LBE-exposed sample), the relative solubility of Bi and Pb in the $Zr_2AlC$ lattice at the formation temperatures, and the relatively stability of oversaturated (in Bi and Pb) MAX phase upon cooling to RT. The major impurity was refined to be $ZrC_{0.63(4)}$ with $a$ = 4.6764(3) Å (Table 1), in very good agreement with [43]. The ZrC competing phase in the $Zr_2AlC$ MAX phase ceramic as well as the $Zr_2AlC$ phase had full C occupancy according to the reported NPD results in the as-sintered state [14].

It is very important to note that the isotropic thermal displacement of 7.6(2) Å$^2$ for the $A$ metal site in the MAX phase is extremely high compared to the other refined thermal displacement values in the two major phases. This value is in excellent agreement with the existence of a high site disorder. In case of high site disorder, individual scatterers are positioned in slightly different positions with respect to the ideal lattice position of the site, and this slight displacement is not coherent between different unit cells. This results in a diffraction pattern similar to that of a high thermal displacement [46]. Such disorder is expected to exist when 3 elements occupy the same site, and this thermal displacement value is another strong indication of that effect.

The refined lattice parameters of the IMC $Zr_2Al_3$ and $ZrAl_3$ minor phases deviate from their previously published values [44,45]. This may result either from the small observed statistics corresponding to these minor phases or from some doping of the Al sites in these phases with LBE. Close examination of such effects is beyond the scope of this paper. Most importantly, due to the very small weight percent of these phases, their exact refinement does not affect any of the main findings and conclusions of the above-mentioned analysis.

## 4. Discussion

*4.1. Zr$_2$AlC/LBE interaction mechanism*

The progressive interaction of the $Zr_2AlC$ ceramic with liquid LBE at 500°C is schematically presented in Fig. 13, summarizing the observed phenomena in a stepwise fashion. In addition to the micrometer scale (macroscopic) changes in the schematic, the atomic changes in the MAX phase lattice are presented in the insets above each step.

In step-I (Fig. 13-I), the polished sample surface is exposed to liquid LBE, which penetrates into the multi-phase ceramic mainly by GB diffusion; both $Zr_2AlC$ and ZrC GBs facilitate LBE ingress. In addition to GBs, high-energy surfaces within the grains, such as SFs and twin boundaries, are also highly susceptible to LBE attack. Despite the LBE ingress, the surface integrity of the MAX phase ceramic is preserved without any apparent material loss.

In step-II (Fig. 13-II), according to EDS results acquired from the $Zr_2AlC$/LBE interaction front, Bi penetrates first into high-energy areas of the MAX phase grains, such as SFs or dislocation walls. This could be mainly attributed to relative size differences between *A*-elements: both Bi (143 pm) and Pb (154 pm) are much larger than Al (118 pm), but since Bi is smaller than Pb, it initiates heavy-atom diffusion into the MAX phase lattice by opening it up, so as to facilitate the ingress of the larger Pb atoms together with more Bi atoms. In turn, Al starts diffusing out of the Zr$_2$AlC grains, dissolving into the liquid LBE due to its high solubility [47,48].

In step-III (Fig. 13-III), as Bi and Pb occupy the SFs along the basal planes, other defects such as dislocation pile-ups become facilitators of the LBE diffusion away from the basal planes. LBE ingress through material defects has also been observed in LBE-exposed 316L austenitic stainless steels, where deformation twin laths were preferentially attacked and ferritized due to the selective leaching of austenite stabilizers (Ni, Mn) [3,4]. The concurrent diffusion of Bi and Pb away from the basal planes will progressively lead to the complete transformation of the Zr$_2$AlC grains into Zr$_2$(Al,Pb,Bi)C solid solution grains (step-IV, Fig. 13-IV).

HRSTEM imaging and EDS mappings showed that the substitution of Al by Pb and Bi in the *A*-layers of the MAX phase lattice is not random, but there is a tendency for atomic ordering. Alternating, intermixed Al:Bi:Pb layers with higher and lower Al contents, create out-of-plane ordered MAX phase domains. The observed antiphase domain boundaries are possibly the result of the multiple points of LBE ingress into the Zr$_2$AlC MAX phase grains (e.g., GBs and SFs), combined with the relatively slower elemental diffusion along the <*c*> direction when compared to elemental diffusion along basal planes. The final microstructure consists of Zr$_2$(Al,Bi,Pb)C MAX phase solid solution and ZrC grains, where the GBs are decorated with Bi, Pb and Al (step-IV, Fig. 13-IV).

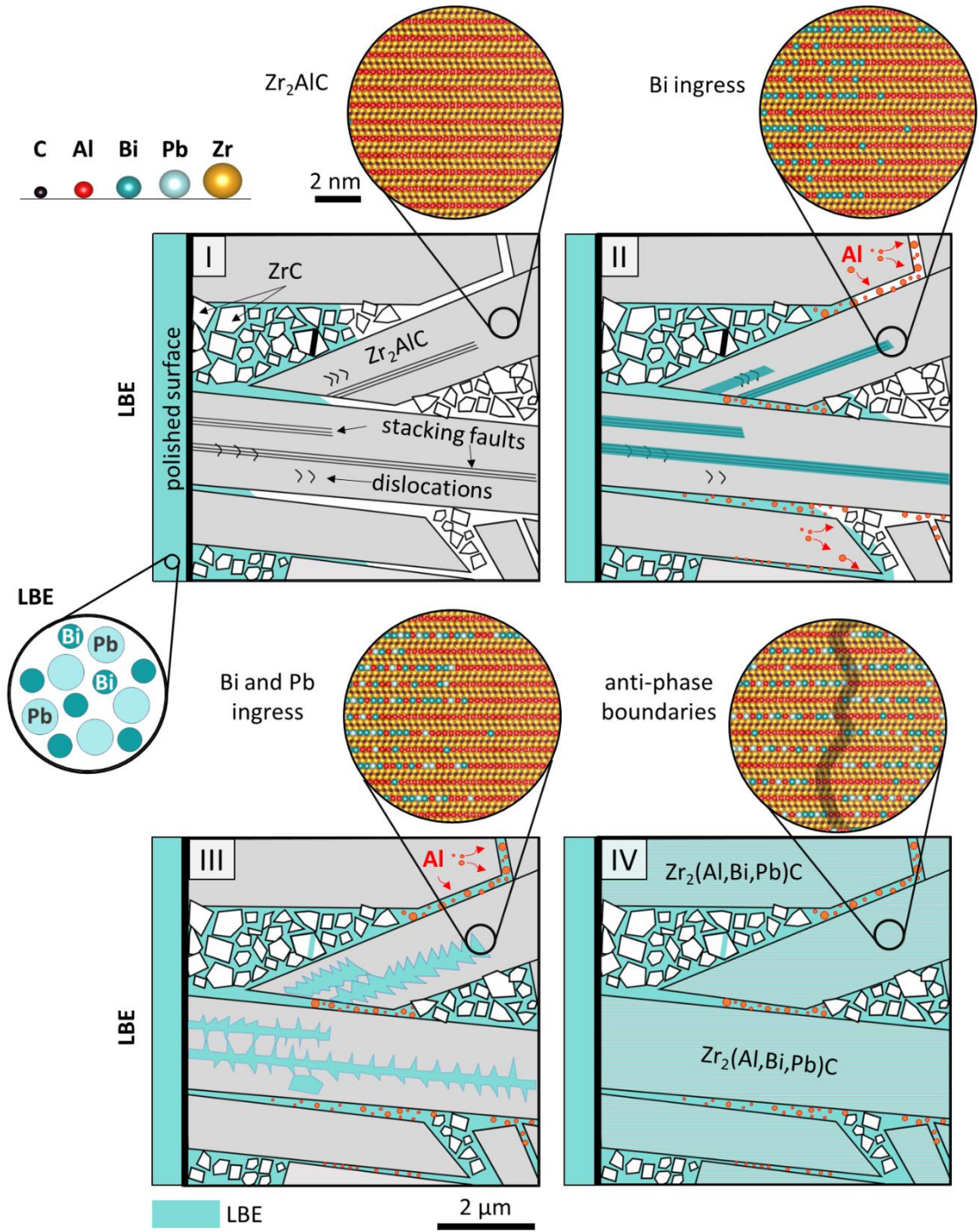

**Fig. 13.** Suggested Zr$_2$AlC/LBE interaction mechanism. <u>Step-I</u>: Bi/Pb attack of Zr$_2$AlC MAX phase GBs. <u>Step-II</u>: The initial Bi ingress into the MAX phase grains through high-energy, defect-rich features, such as SFs and dislocation pile-ups leads to Al outward diffusion. <u>Step-III</u>: Bi/Pb occupation of basal SFs and further diffusion in other directions. The Bi/Pb ingress into the *A*-layers of Zr$_2$AlC is characterized by an ordering tendency, resulting in alternating Al-rich and Bi/Pb-rich layers. <u>Step-IV</u>: formation of nano-domains of the Zr$_2$(Al,Bi,Pb)C solid solution with out-of-plane ordering tendency.

*4.2. ZrC/LBE interaction mechanism*

ZrC grains were observed to be chemically compatible with liquid LBE. No Bi or Pb were ever measured inside the ZrC grains. However, similar to the high-energy, defect sites (GBs, SFs) in the $Zr_2AlC$ MAX phase grains, twins provided a path for preferential Bi and Pb ingress, as shown in Fig. 8b, and most of the ZrC grains examined by TEM in this work had twins. Similar to other transition metal carbide rock salt structures, ZrC exhibits a wide range of non-stoichiometric $ZrC_x$ compounds, with $0.6 < x < 1.0$. These C-substoichiometric binary carbides are considered to be stable in their ordered state at low temperatures. High-temperature sintering usually leads to disordered C vacancies, while the formation of ordered phases, such as $Zr_2C$, requires long-term annealing [49,50]. In this respect, the LBE exposure (500°C, 1000 h) herein discussed can be considered as a long annealing treatment. Recently, vacancy-ordered $ZrC_{0.61}$ had been synthesized by spark plasma sintering and its twinning behavior was investigated [34]. Compared to the high stacking fault energy stoichiometric ZrC, the vacancy-ordered $ZrC_{0.61}$ has substantially lower twinning energies and more available twin types [51]. Although no evidence for ordered ZrC superlattice diffractions in XRD or SAED patterns was found in this study (these techniques have also limited sensitivity to C near heavy Zr atoms), it is possible that these diffractions were very weak to be recorded; therefore, one cannot fully exclude the possibility that a vacancy-ordered $ZrC_x$ might be forming during the LBE exposure. More work is needed to quantify the composition of the $ZrC_x$ phase, which is beyond the scope of this paper. It is worthwhile mentioning that the $ZrC_x$ phase in the reactive hot pressed $Zr_2(Al,Bi,Pb)C$ had a stoichiometry of $x = 0.63$, as measured by NPD, but no ordering was detected. The ZrC that formed during the synthesis of the $Zr_2AlC$-based ceramic [14], which was tested in LBE was initially stoichiometric. The effect of the tendency to form C vacancy-ordered $ZrC_x$ on MAX phase synthesis is unknown, yet it is possible that a similar ordering tendency might be present or a disordered state might be stabilized within the octahedral $Zr_6C$ layers in the relatively more complex MAX phase unit cell. In $Zr_3Al_3C_5$ (i.e., a Zr-Al-C compound with similar structure to the MAX phases, but containing three *A*-layers instead of one), twinned ZrC inclusions were observed within the $Zr_3Al_3C_5$ grains [52]. These twins had a thin Al platelet at the ZrC twin boundary [53], similar to what was observed in this study, which detected Al-Bi-Pb at the twin boundaries (Fig. 8c). The presence of Al in the ZrC twin boundaries was attributed to the reduction of the twin boundary energy by the segregation of impurities such as Al and Si [51,53]. TEM investigations have indicated that Al decreases the ZrC twin boundary energy, possibly forming ternary Zr-Al-C phases by intercalating Al (or $Al_3C_2$) in the ZrC twin boundaries. The findings of this work with respect to twin boundary decoration in ZrC by Al, Pb and Bi (Fig. 8b) might indicate MAX phase formation at these twin boundaries. However, more HRTEM and analytical work is needed to confirm this hypothesis.

*4.3. The chemical instability of $Zr_2AlC$ in liquid LBE*

One of the first steps of the LBE corrosion mechanism is the outward diffusion of Al from the $Zr_2AlC$ grains. The outward diffusion of Al has been commonly reported in other MAX phases. Crack healing in $Ti_2AlC$ and $Cr_2AlC$, for example, was reported to occur by the outward diffusion of Al from these MAX phases to form $Al_2O_3$ in oxidizing environments [54,55]. In $Cu/Ti_3AlC_2$ composites, Al outward diffusion through basal planes above 950°C was observed, forming a Cu(Al) solid solution and leaving substoichiometric $TiC_{0.67}$ behind [56].

Outward diffusion of the *A*-element was observed in corrosion studies as well. Corrosion tests in liquid sodium (Na) at 750°C for 168 h [57] showed the outward Al diffusion from the $Ti_3AlC_2$ grain bulk to the GBs, leaving Al-substoichiometric grains. From the GBs, Al diffused into the liquid Na. On the contrary, no variation in the Ti/Al ratio in $Ti_2AlC$ or outward Al diffusion was reported. It should be noted that Al and Na have almost zero solubility in each other [58]. The $Ti_3SiC_2$, $Cr_2AlC$ and $Ti_2AlC$ compounds were resistant to Na corrosion, while $Ti_3AlC_2$ interacted with Na, resulting in compositional and mechanical property changes due to Al loss and GB attack. Corrosion tests performed on $Ti_3SiC_2$, $Ti_3AlC_2$, $Ti_2AlC$ and $Cr_2AlC$ in simulated primary water for light water nuclear reactors showed the outward diffusion of Al and Si in Ti-based MAX phases, while O and Fe in the water diffused inwards to form $TiO_2$ and $FeTiO_3$ [59].

Palladium (Pd) and $Ti_2AlC$, $Cr_2AlC$, $Ti_3AlC_2$ or $Ti_3SiC_2$ diffusion couples were annealed at 900°C for 10 h to test these MAX phases as diffusion barriers against fission products, such as Pd, in TRISO fuel designs [60]. Similar to other interaction tests, Al outward diffusion occurred in the investigated MAX phases, forming MAX phase solid solutions by the ingress of Pd. $Ti_2AlC$ performed best against Pd attack, whereas $Ti_3SiC_2$ was the worst.

As a general rule, it was previously suggested that the MAX phases would not react with Pb [61] or Pb-Bi [6] alloys, if no intermetallic formed or the solid solubility of the *A*-group element in this alloy was negligible. Among the Zr-Al-C MAX phase constituents, Al is the most soluble element in liquid Pb (0.038 mass% at 500°C) [47] and liquid Bi (2 mass% at 600°C) [48]; hence, the $Zr_2AlC$ MAX phase could indeed undergo selective leaching of Al in contact with liquid LBE. The Zr solubility in liquid LBE is higher than that of Cr and Fe, but lower than that of Ni [62]. Selective leaching of Ni and Mn from austenitic stainless steels exposed to oxygen-poor, liquid LBE at 500°C, for example, has been reported to result in steel ferritisation [3]. Moreover, the Zr solubility in liquid LBE is comparable to the solubility of Ti at 500°C [63]; however, Ti-based MAX phases have been observed to behave well in liquid LBE [5,6,13]. It is thought that the complex carbide structure and strong Zr(Ti)-C bonds in the $Zr_2AlC$ MAX phase might have prevented the dissolution of Zr in liquid LBE. This agrees with the fact that ZrC was also very resistant to LBE attack, therefore, Zr is less likely to be leached out of the MAX phase when compared to Al. Moreover, no GB enrichment in Zr was detected in this study. Lastly, C has a very low solubility in LBE ($0.12 \times 10^{-2}$ mass% at 900°C) [64], making its dissolution in liquid LBE an unlikely event. Both experimental findings and previously published work suggest that the Al outward diffusion plays a critical role in the $Zr_2AlC$/LBE interaction mechanism.

The outward diffusion of the *A*-element and its substitution with another *A'*-element is used as a synthesis technique for new (ternary) MAX phases and MAX phase solid solutions. The synthesis process mostly consists of capping a thin film MAX phase with various *A*-elements and heating it at relatively low temperatures (in the 300-670°C range, much lower than bulk synthesis temperatures) for relatively long durations (6-24 h). The following MAX phases were synthesised by replacing the *A*-element: $Ti_3AuC_2$, $Ti_3Au_2C_2$ and $Ti_3IrC_2$ from $Ti_3SiC_2$ [65]; $Ti_2Au_2C$ and $Ti_3Au_2C_2$ from $Ti_2AlC$ and $Ti_3AlC_2$ [66]; $Mo_2AuC$ and $Mo_2(Au_{1-x},Ga_x)_2C$ from $Mo_2GaC$ and $Mo_2Ga_2C$ [67]; $(Cr_{0.5},Mn_{0.5})_2AuC$ from $(Cr_{0.5},Mn_{0.5})_2GaC$ [68]; and $Mo_2(Fe,Ga,Au)C$ (with Fe:Ga:Au 3:2:1) from $Mo_2GaC$ thin films [68]. In $Mo_2(Au_{1-x},Ga_x)_2C$, one-dimensional in-plane ordering was observed with Au:Ga ≈ 9:1 in the *A*-layer, where one lighter Ga ion was followed by nine heavier Au atoms in the [11$\bar{2}$0] direction [67].

These prior literature studies indicate that the in-situ formation of a new MAX phase solid solution containing Pb and/or Bi is highly likely during the Zr$_2$AlC exposure to liquid LBE, since substitution of *A*-elements is commonly observed. While binary phase diagrams, for both Al-Bi and Al-Pb exist, no ternary or quaternary diagrams in the Zr-Al-Pb-Bi system have been reported to the authors' knowledge. Both Bi and Pb are among the group 13-15 elements in the periodic table and are among the group of 'MAX phase forming' *A*-elements. In fact, Zr$_2$PbC [69,70] and Zr$_2$(Al$_{0.42}$,Bi$_{0.58}$)C [39] have already been experimentally synthesized. In addition, random mixing energies have been calculated for various *A*-elements in Zr$_2$(Al,*A*)C MAX phases, and mixing Al with both Bi and Pb was found to be energetically favorable, forming MAX phase solid solutions [71].

DFT calculations can be used as a guide to assess the possibility of vacancy formation in Zr$_2$AlC, showing that the relative vacancy formation energies are: $E_{vac}$ (Zr) > $E_{vac}$ (C) > $E_{vac}$ (Al) [72]. Al vacancy formation is clearly more favourable, but becomes more difficult under Zr- and/or Al-rich conditions. Under these conditions, C vacancies become the most stable. The occasional measured Al-over/substoichiometry within the Zr$_2$AlC grains in this study might indicate that, depending on the local Al concentration, C vacancies might be more favourable in the annealed or LBE-exposed Zr$_2$AlC. In terms of interstitial defects, Al interstitials are the easiest to form in Zr$_2$AlC, irrespective of chemical potentials, and they prefer to relax in the existing Al-layer. To support this, the bond lengths and interplanar spacings of fcc Al were compared with those of Zr$_2$AlC by Shah et al. [72], comparing the Al interplanar spacing of $d_{Al\{110\}}$ and $d_{Al\{111\}}$ with those of the Zr$_2$AlC MAX phase $d_{Al-Al}$ and $d_{Al-M}$, respectively. As the $d_{Al-M}$ > $d_{Al\{111\}}$, it was shown that Al atoms in a Zr$_2$AlC unit cell are in tension along the <*c*> direction when compared to its natural fcc state. This comparative percent tensile (or compressive) strain values along the <*c*> direction can be calculated as follows and are given in Table 2:

$$Lattice\ strain\ (\%) = -(\frac{d_{Al\{111\}} - d_{Al-M}}{d_{Al\{111\}}}) \times 100 \qquad (4)$$

Due to this tension, Al interstitials are more likely to form and remain in the Al-layer so as to compensate the stress. The tension around the *A*-layer in the Zr$_2$AlC MAX phase and its concomitant sink function for Al interstitials might also indicate why Bi/Pb atoms attacked Zr$_2$AlC and other Zr-rich Al based MAX phases, while many other MAX phases did not interact at all [13]. Using a similar approach, one can assume the initial Bi and Pb ingress to disturb the *A*-layer and substitute for the host *A*-atoms, knowing they will replace Al in Zr$_2$AlC. The nature of the surrounding stress field around Al atoms might be important for potential Pb/Bi ingress and concomitant creation of Al (or *A*) interstitial defects. Tension in the bonds might facilitate the relatively large Bi/Pb atoms to enter the Zr$_2$AlC lattice in order to relax the MAX phase unit cell. For comparison, the bond lengths and interplanar spacings in Zr$_2$AlC and other MAX phases tested in the same LBE exposure experiment are summarised in Table 2. Strain values > 0 indicate tensile strain. Note that only in those MAX phases reported to react and to form solid solutions with Bi/Pb, the stress field around the Al atoms was in tension when compared to the natural state of fcc Al.

**Table 2.** Comparison of bond lengths and interplanar spacings for MAX phases tested in LBE [13] and those of related *A*-elements. The ones reported to form in-situ solid solutions with LBE are presented in **bold**. References with 'mp' code are retrieved from [76].

| MAX phases | $d_{Al-Al}$ (Å) | $d_{Al-M}$ (Å) | Strain (%) | Ref. | A-element | $d_{\{110\}}$ (Å) | $d_{\{111\}}$ (Å) | Ref. |
|---|---|---|---|---|---|---|---|---|
| $Zr_2AlC$ | 3.33 | 2.386 | **2.38** | [72] | | | | |
| $Zr_2AlC$ | 3.324 | 2.334 | **0.17** | [14] | | | | |
| $Ti_2AlC$ | 3.069 | 2.288 | -1.82 | mp-12990 | | | | |
| $Ti_3AlC_2$ | 3.081 | 2.302 | -1.20 | mp-3747 | | | | |
| $Nb_4AlC_3$ | 3.16 | 2.261 | -2.98 | mp-3103 | | | | |
| $Nb_2AlC$ | 3.134 | 2.237 | -4.01 | mp-996162 | Al | 2.856 | 2.33 | mp-134 |
| $Cr_2AlC$ | 2.843 | 2.1 | -9.87 | mp-9956 | | | | |
| $(Nb_{0.85},Zr_{0.15})_4AlC_3$ | 3.155 | 2.21 | -5.17 | [77] | | | | |
| $(Zr_{0.8},Ti_{0.2})_2AlC$ | 3.26 | 2.397 | **2.85** | [78] | | | | |
| $(Zr_{0.8},Ti_{0.2})_3AlC_2$ | 3.29 | 2.386 | **2.40** | [78] | | | | |
| $Zr_3AlC_2$ | 3.333 | 2.414 | **3.61** | [79] | | | | |
| $Ti_3SiC_2$ | 3.076 | 2.042 | -35.33 | mp-5659 | Si | 3.867 | 3.157 | mp-149 |
| $(Zr_{0.5},Ti_{0.5})_2(Al_{0.5},Sn_{0.5})C$ | 3.234 | 2.337 | -3.67 | [73] | Sn | 2.523 | 4.129 | mp-117 |

A similar approach would be to calculate lattice distortions in the $Zr_2(Al,Bi,Pb)C$ MAX phase solid solution formed during the exposure of the $Zr_2AlC$ MAX phase to liquid LBE. The successful implementation of this approach depends on the accuracy of the *a*, *c*, and $z_M$ (*z*-coordinate of the *M* (Zr) atom) parameters, all of which are changing depending on the exact MAX phase solid solution composition and are difficult to determine for local and limited LBE-affected volume. Still, calculating the distortions [41] in the $Zr_2(Al,Bi,Pb)C$ MAX phase solid solution, using the NPD data acquired from the hot pressed $Zr_2(Al,Bi,Pb)C$ ceramic, results in a reduced trigonal prism distortion of 1.080 ($Zr_2AlC$ = 1.101 [42]) and an increased octahedral distortion of 1.1103 ($Zr_2AlC$ = 1.034 [42]). Recent studies showed that aiming to reduce the trigonal prism distortion by making double solid solution MAX phases (i.e., solid solutions on both *M* and *A* sites of the 'host' MAX phase compounds) enhanced the phase purity of the produced ceramics [42,73]. Ingress of Bi and Pb seems to lower the trigonal prism distortions in the new $Zr_2(Al,Bi,Pb)C$ MAX phase solid solution in comparison with the pristine $Zr_2AlC$.

More detailed work is needed to confirm the effect of distortions on corrosion resistance, yet it might be safe to assume that, depending on the atomic size of the exposure medium, a proper selection of the *A*-element in the pristine MAX phase might facilitate the compression or tension, thus affecting the tendency towards in-situ MAX phase solid solution formation.

Another factor to consider when assessing the compatibility of $Zr_2AlC$ with liquid LBE is the delaminations observed after the long-term (several months to 1 year) storage of LBE-exposed $Zr_2AlC$ at RT. These delaminations are believed to have resulted from a low-temperature LBE transformation. At 123.5°C, a eutectic reaction L → γ+β occurs, where the γ-phase is a solid solution of Pb in Bi (Bi with 0.4% Pb) and the β-phase is an IMC (42 wt% Bi at 123.5°C, and 33.8 wt% Bi at 0°C). According to the Pb-Bi phase diagram, after cooling to RT, the close-packed β-phase (hcp; density of 11.17 g/cm$^3$) transforms into the less dense γ-phase (rhombohedral; density 9.747 g/cm$^3$). Upon fast cooling, equilibrium conditions are not achieved, and the transformation of the dense β-phase to the less dense γ-phase is incomplete, continuing slowly during storage [4,74]. Consequently, LBE-occupied regions (GBs or twin boundaries) are likely to degrade in time due to the volume expansion accompanying this transformation, resulting in microcracks and delaminations. Even though this phenomenon is only expected to occur after the complete shutdown of a nuclear reactor [4], it should be taken into account during material characterization studies, as it changes the material microstructure in time and endangers the sample integrity, especially for thin FIB foils.

Overall, the composition of the $Zr_2$(Al,Bi,Pb)C MAX phase solid solution varied, showing an Al:Bi:Pb ratio between 3:2:1 and 6:2:1. This ratio is thought to relate to the Bi and Pb solubility in the MAX phase *A*-layer at the exposure temperature (500°C) and at RT, and the extent of Bi/Pb ingress at the specific locations examined in this work. DFT calculations confirmed that this range of compositions in the $Zr_2$(Al,Bi,Pb)C solid solution is indeed favourable (Fig. 12).

One may add that the in-situ formation of the $Zr_2$(Al,Pb,Bi)C MAX phase solid solution resulting from the exposure of $Zr_2$AlC to liquid LBE is accompanied by a reduction in neutron absorption for both thermal and fast neutron spectra, as compared to $Zr_2$AlC [75]. Therefore, the in-situ formation of $Zr_2$(Al,Bi,Pb)C essentially translates into the progressive reduction of any possible neutronic penalty stemming from the deposition of $Zr_2$AlC coatings on Gen-IV LFR fuel clads. $Zr_2$PbC and $Zr_2$(Pb,Bi)C have been previously proposed as best candidate MAX phase materials from the neutronics point of view [75]. Using $Zr_2$AlC coatings for the protection of Gen-IV LFR fuel clads from liquid metal attack becomes appealing only when the deposition temperature of (phase-pure) coatings is sufficiently low, so as to avoid altering the bulk properties of the substrate fuel clad. Moreover, the $Zr_2$AlC coatings should maintain their structural integrity and attractive properties (thermal conductivity, damage tolerance) during long-term service in contact with the HLM coolant.

*4.4. The superlattice in annealed $Zr_2$AlC*

Regarding the superlattice observed in LBE-unaffected $Zr_2$AlC grains after prolonged annealing as well as in vacuum-annealed $Zr_2$AlC ceramics, very limited work is available in literature. To the authors' knowledge, apart from the i-MAXs, the only other reported MAX phases with superlattices are the vacancy-ordered $V_{12}Al_3C_8$ [$V_4AlC_{(3-y)}$ (y ≈ 0.31)] [36] and $Nb_{12}Al_3C_8$ [$Nb_4AlC_{(3-z)}$ (z ≈ 0.3)] [35]. Both superlattice structures had an $a' = a\sqrt{3}$ sized unit cell without change in *c* lattice parameter. Due to the ordered C vacancies, the *M*-atoms in these phases were slightly misplaced in directions away from the vacancy sites, which affected the diffraction intensities.

For the ordered $Nb_4AlC_{(3-y)}$ phase, 120 s of electron beam imaging were reported to be enough to make the superlattice reflections disappear and transform into a disordered phase [35]. For the $V_{12}Al_3C_8$ phase, no additional XRD superlattice peaks were observed when the sample was in powder form, and the authors reported that working with single crystals was essential for identifying the structure by XRD [36]. Similar difficulties also hindered the determination of the exact superlattice structure observed in this study. The tendency of C vacancy ordering in the MAX phases can be linked with the intrinsic tendency of the respective *M*-C compounds (*M* = Zr, Ti, V, Nb) towards superlattice formation. For example, ZrC exhibits substoichiometric superlattice structures at low temperatures [34]. To the authors' knowledge, no study exists on the stability of the MAX phases under prolonged annealing treatments. The intrinsic lattice distortions in the $Zr_2AlC$ unit cell might also affect the tendency towards superlattice formation in this MAX phase structure.

The relation between the superlattice structure generated upon annealing $Zr_2AlC$ at 500°C for 1000 h and the $Zr_2AlC$/LBE interaction mechanism is unclear at the moment. However, since the in-situ formed $Zr_2(Al,Bi,Pb)C$ MAX phase solid solution probably does not retain the superlattice structure (no superlattice reflections were observed in the SAED patterns), it is possible that any vacancy or anti-site defects generated in an orderly manner during annealing are recoverable by the new Bi- and Pb-containing solid solution or become disordered again. The NPD analysis on hot pressed $Zr_2(Al,Bi,Pb)C$ revealed a C stoichiometry of 0.81, indicating that some amount of C vacancies might remain as favourable in this MAX phase solid solution.

*4.5. Proposed approach for mitigating the $Zr_2AlC$/LBE interaction*

The examinations of the LBE-exposed $Zr_2AlC$-based ceramics revealed the presence of a thin (Zr,Al)-containing oxide scale, which did not appear to be protective against LBE attack, as it was detected in both LBE-affected and unaffected areas (Fig. 5). No systematic study on the protectiveness of the oxide layer as function of its thickness was performed in this work. However, it is thought that carbides can be effective barriers against LBE ingress, at least for a (temperature-dependent) period of time, since it was found that ZrC grain clusters were practically impervious to LBE penetration (Fig. 5a), while no such ZrC layer was present in the LBE-affected area (Fig. 5b). More detailed examinations as well as additional LBE exposures are needed to assess the actual protectiveness of ZrC barrier layers and further dedicated investigations are currently ongoing.

## 5. Conclusions

In this work, a detailed characterization of LBE-exposed $Zr_2AlC$-based MAX phase ceramics was performed by SEM, EDS, (S)TEM, XRD, APT and NPD. The $Zr_2AlC$ ceramic had been exposed to oxygen-poor ($C_O \leq 2.2 \times 10^{-10}$ mass%), static liquid LBE at 500°C for 1000 h. The main findings of this study may be summarized as follows:

1. Local interaction was observed in LBE-exposed $Zr_2AlC$. In the affected areas, Bi and Pb atoms diffuse first along $Zr_2AlC$ and ZrC grain boundaries (GBs), before diffusing into the bulk of the MAX phase grains. Surface integrity was preserved and no material loss was observed. In the non-affected areas, ZrC grain clusters were found at the sample surface, inhibiting the interaction with LBE.
2. Al outward diffusion from the $Zr_2AlC$ MAX phase grains to GBs, followed by Bi/Pb ingress into the $Zr_2AlC$ grains, resulted in the in-situ formation of a new $Zr_2(Al,Bi,Pb)C$ MAX phase solid solution. Initially, only Bi was found in partially affected $Zr_2AlC$ grains. Bi/Pb atoms initially penetrated into the basal planes, potentially along stacking faults (SFs), or defect-rich areas in the MAX phase grains, to finally spread throughout the whole $Zr_2AlC$ grains.
3. The $Zr_2(Al,Bi,Pb)C$ MAX phase solid solution showed a tendency for out-of-plane ordering in the mixed Al:Bi:Pb *A*-layers, with alternating high and low Al contents, as confirmed by DFT calculations, thus creating ordered domains separated by antiphase boundaries.
4. The LBE exposure/annealing at 500°C for 1000 h resulted in a superlattice structure in the non-LBE affected $Zr_2AlC$ MAX phase, potentially due to ordered C vacancies and/or Al anti-site defects. Understanding the stability of $Zr_2AlC$ under long-term annealing requires further work.
5. NPD and EDS performed on reactive hot pressed $Zr_2(Al,Bi,Pb)C$ MAX phase confirmed that this phase can also be synthesized with conventional processing techniques. The lattice parameters of the new $Zr_2(Al,Bi,Pb)C$ solid solution were refined as $a$ = 3.3472(2) Å and $c$ = 14.5401(8) Å.

The findings of this work point towards two promising directions of further research: (1) ZrC-based diffusion barriers on the surface of $Zr_2AlC$ ceramics might effectively mitigate the $Zr_2AlC$/LBE interaction for a sufficiently long period of time (e.g., the lifetime of the substrate fuel clad), and (2) the in-situ formed $Zr_2(Al,Bi,Pb)C$ solid solution is more stable in contact with liquid LBE than $Zr_2AlC$ and with an even smaller neutron absorption, making it an interesting compound for Gen-IV LFR fuel cladding applications. More work is needed to improve the phase purity of ceramics/coatings made of this solid solution and to assess its full potential.

**Data availability**

The raw/processed data required to reproduce these findings cannot be shared at this time, as the data also form part of an ongoing study.

**Acknowledgements**

B.T. acknowledges the financial support of the SCK•CEN Academy for Nuclear Science and Technology. This research was partly funded by the European Atomic Energy Community's (Euratom) Seventh Framework Programme FP7/2007-2013 under Grant Agreement No. 604862 (FP7 MatISSE), the MYRRHA project (SCK•CEN, Belgium), as well as by the Euratom research and training programme 2014-2018 under Grant Agreement No. 740415 (H2020 IL TROVATORE). The performed research falls within the framework of the EERA (European Energy Research Alliance) Joint Programme on Nuclear Materials (JPNM). The authors gratefully acknowledge the Hercules Foundation for Project AKUL/1319 (CombiS(T)EM)) and the Knut and Alice Wallenberg (KAW) foundation. The calculations were carried out using supercomputer resources provided by the Swedish National Infrastructure for Computing (SNIC) at the High Performance Computing Center North (HPC2N) and the PDC Center for High Performance Computing. E.N.C thanks Offir Ozeri for his help in NPD data acquiring.

**Appendix: Supplementary data**

The following are Supplementary data to this article:
Supplementary_data_LBEZr2AlC.docs